\documentstyle[epsf]{l-aa}

\def\ifigx#1#2{\epsfxsize=#1 \centering{\mbox{\epsfbox{#2}}}}
\def\kms{~km\,s$^{-1}$}
\def\coravel{{\small CORAVEL}}
\def\hd{{\small HD}}
\def\bd{{\small BD}}

\begin{document}

\thesaurus{08.02.4 -- 08.03.2 -- 08.12.1} 

\title{A {\large\bf CORAVEL} radial-velocity monitoring of giant Ba
and S~stars: spectroscopic orbits and intrinsic
variations (I)\thanks{Based on observations obtained at the Haute-Provence
Observatory (France) and at the European Southern Observatory (ESO, La Silla, 
Chile)}}
\author{S.~Udry$^1$, A.~Jorissen$^2$, M.~Mayor$^1$ and S.~Van Eck$^{1,2}$}
\offprints{S.~Udry }
\institute{$^1$Observatoire de Gen\`eve, CH-1290 Sauverny, Switzerland \\
$^2$Institut d'Astronomie et d'Astrophysique, Universit\'e 
Libre de Bruxelles, C.P.226, Bvd du Triomphe, B-1050 Bruxelles,
Belgium }
\date{Received date; accepted date}
\maketitle
\markboth{Spectroscopic orbits of giant Ba and S stars}{Spectroscopic
orbits of giant Ba and S stars}

\begin{abstract}
With the aim of deriving the binary frequency among Ba and S stars, 56
new spectroscopic orbits (46 and 10, respectively) have been derived
for these chemically-peculiar red giants monitored with the \coravel\
spectrometers.  These orbits are presented in this paper (38 orbits)
and in a companion paper (Udry et~al. 1998, Paper~II; 18~orbits). The
results for 12 additional long-period binary stars (6 and 6,
respectively), for which only minimum periods (generally exceeding
10\,y) can be derived, are also presented here (10) and in Paper~II
(2). The global analysis of this material, with a few supplementary
orbits from the literature, is presented in Jorissen et al. (1998).

For the subsample of Mira S, SC and (Tc-poor) C stars showing
intrinsic radial-velocity variations due to atmospheric phenomena,
orbital solutions (when available) have been retained if the velocity
and photometric periods are different (3 stars). However, it is
emphasized that these orbit determinations are still tentative.  Three
stars have been found with radial-velocity variations synchronous with
the light variations. {\sl Pseudo-}orbital solutions have been derived
for those stars. In the case of RZ Peg, a line-doubling phenomenon is
observed near maximum light, and probably reflects the shock wave
propagating through the photosphere.

\end{abstract}

\section{Introduction}

Several samples of chemically-peculiar red giants (PRG) were monitored
in radial velocity with the aim of deriving their binary frequency.
The spectroscopic-binary orbits presented here and in a companion
paper (Udry et al. 1998, Paper~II) belong to a systematic survey of
barium and Tc-poor S~stars undertaken to gain insight into the
formation process of these stars.

Barium stars and their Pop.~II counterparts, the CH~stars, are not
evolved enough to have produced on their own the observed
overabundances of elements heavier than Fe (e.g. Lambert 1985),
usually associated with the nucleosynthesis occurring during
He-burning thermal pulses on the asymptotic giant branch (AGB). The
binary nature of these stars accounts for the observed chemical
peculiarities through mass transfer across the binary system (McClure
et al. 1980; McClure 1983). The exact mass-transfer mode is still a
matter of debate but is turning towards a solution involving
Roche-lobe overflow or wind accretion, depending on the orbital period
(Han et al. 1995).

Contrarily to Tc-rich S~stars which are able to synthesize heavy
elements on their own, Tc-poor S stars are believed to be the cool
descendants of barium stars and thus also owe their chemical
peculiarities to binarity (Jorissen \& Mayor 1992; Ake 1997; Van~Eck
et al. 1997).

The 46 new (43) or updated (3) spectroscopic orbits of barium stars
(+~6 Ba stars with $P_{\rm min}$ determinations) and the 10 new orbits
of S stars (+~6 binary S stars with $P_{\rm min}$ determinations)
reported in this paper and in Paper~II, combined with a few other
newly available orbits (see references in Jorissen et al. 1998),
provide an unequalled sample to constrain the binary evolution
channels relevant for barium stars, as identified by Han et al. (1995)
using only 17 barium-star orbits. It also allows us to rediscuss the
evolutionary link between barium and Tc-poor S~stars.  This is done in
a companion paper (Jorissen et al. 1998).

In order not to bias the S-star sample towards low luminosities, a
small sample of S Miras, SC and Tc-poor C stars were also
monitored. However, the radial-velocity jitter associated with
atmospheric motions (envelope pulsation, large convective cells,
shocks) induces non-orbital radial-velocity variations that limit our
ability to detect binaries and, {\it a fortiori}, to derive their
orbital elements, at a given level of measurement
precision. Illustrative examples will be discussed at the end of the
paper.

This paper is organized as follows. Section~\ref{sect2} describes the
stellar samples. The observations are briefly discussed and a short
statistical overview of the survey is given in
Sect.~\ref{sect3}. Section~\ref{sect4} provides the orbital elements
and the radial-velocity curves. Some stars are also commented
individually.  Mira S, SC and C stars with radial-velocity variations
associated or not with an orbital motion are discussed in
Sect.~\ref{sect5}.

\section{CORAVEL star samples and binary content}
\label{sect2}

The star samples considered in this paper were observed with the
\coravel\ spectro-velocimeters (Baranne et al. 1979) installed on the
1-m Swiss telescope at the Haute-Provence Observatory (France) and on
the 1.54-m Danish telescope at the European Southern Observatory (La
Silla, Chile). The criteria defining these various samples, their
monitoring history and their binary content are described in the
following subsections. For a detailed discussion of the binary
frequency in the corresponding PRG family as a whole, we refer to the
analysis paper (Jorissen et al. 1998).

\begin{table}[th!]
\begin{flushleft}
\caption[]{The sample of barium stars with strong anomalies (Ba4 or
Ba5 on the scale defined by Warner 1965) monitored with \coravel. The
spectral type, visual magnitude and $B-V$ index are from L\"u et
al. (1983) or, if not available, from L\"u (1991).  The last column
refers to a note at the end of the table providing the reference where
the orbital elements are to be found }
\label{tab1} 
\begin{tabular}{lccccc}
\hline
HD/DM/.. & Sp. & Ba & $V$ & $B-V$ & orbit \\ 
\hline
5424   & K1 & 4 & 9.48 & 1.14 & 1 \\
19014  & K4 & 5 & 8.18 & 1.69 & 0 \\
20394  & K0 & 4 & 8.72 & 1.09 & 6 \\
24035  & K4 & 4 & 8.48 & 1.24 & 1 \\
36598  & K2 & 4 & 8.02 & 1.31 & 1 \\
42537  & K4 & 5 & 8.93 & 1.85 & 1 \\
43389  & K2 & 5 & 8.32 & 1.48 & 2 \\
44896  & K3 & 5 & 7.26 & 1.58 & 2 \\
46407$^a$  &K0  & 3 & 6.28 & 1.12  & 2 \\
50082  & K0 & 4 & 7.42 & 1.02 & 2 \\
60197  & K3 & 5 & 7.74 & 1.69 & 1 \\
84678  & K2 & 4 & 9.00 & 1.46 & 1 \\
88562  & K2 & 4 & 8.54 & 1.42 & 1 \\
92626  & K0 & 5 & 7.10 & 1.34 & 2 \\
100503 & K3 & 5 & 8.74 & 1.69 & 1 \\
107541 & K0 & 4 & 9.39 & 1.06 & 2 \\
120620 & K0 & 4 & 9.62 & 1.10 & 1 \\
121447$^c$ & K7 & 5 & 7.85 & 1.75 & 7 \\
123949 & K6 & 4 & 8.73 & 1.37 & 1 \\
154430 & K2 & 4 & 8.74 & 1.56 & 1 \\
196445 & K2 & 4 & 9.08 & 1.42 & 1 \\
201657 & K1 & 4 & 8.04 & 1.22 & 2 \\
201824 & K0 & 4 & 8.91 & 1.08 & 6 \\
211594 & K0 & 4 & 8.11 & 1.16 & 2 \\
211954 & K2 & 5 &10.21 & 1.33 & 1 \\
+38$^\circ$\ 118
       & K2 & 5 & 8.88 & 1.50 & 1 \\
$-42^\circ$2048
       & K2 & 4 & 9.30 & 1.41 & 1 \\
$-64^\circ 4333^d$
       & K0 & 4 & 9.60 & 1.20 & 1 \\
L\"u 163 
       & G5 & 5 &10.90 & 1.18 & 1 \\
\hline
\end{tabular}\\
$a$) {\small DAO}+\coravel\ measurements \\ 
$c$) Two classifications: Ba5 or S0. Appears in both tables\\ 
$d$) CpD \\
References to orbital elements: 0: Jorissen et al. (1998); 1:~this
paper; 2:~Udry et al. (1998, Paper~II); 6:~Griffin et al. (1996);
7:~Jorissen et al. (1995)
\end{flushleft}
\end{table}

\subsection{Strong barium stars}

This sample includes the 28 barium stars with a strong anomaly (Ba4 or
Ba5 on the scale defined by Warner 1965) from the list of L\"u et
al. (1983), not already monitored by McClure at the Dominion
Astrophysical Observatory (DAO), plus the special star \hd~46407 (see
Sect.~\ref{sect24}). These stars are presented in
Table~\ref{tab1}. The \coravel\ monitoring of that sample started in
1984 (preliminary results were presented in Jorissen \& Mayor 1988).
All the stars show radial-velocity variations due to binary motion
except \hd~19014 for which the situation is still unclear.  Among the
new spectroscopic orbits of strong barium stars, 18 are presented in
Sect.~\ref{sect4}. The others can be found in Paper~II (8), in Griffin
et al. (1996; 2) and in Jorissen et al. (1995; 1). The corresponding
references are indicated in the last column of the table.

\subsection{Mild barium stars}

A random selection of 33 stars with a mild barium anomaly (Ba$<$1,
Ba1, Ba2 on the Warner scale, from the list of L\"u et al. 1983) has
been monitored since 1988 for comparison. A few mild barium stars
observed by the Marseille team (see Paper~II) for a different purpose
were included in our initial sample later on.  The global sample, also
including 3 DAO stars (see next section), is presented in
Table~\ref{tab2}.

\begin{table}[th!]
\begin{flushleft}
\caption[]{The sample of barium stars with mild anomalies (Ba$<$1, 
Ba1, Ba2 on the scale defined by Warner 1965) monitored with
\coravel. The spectral type, visual magnitude and $B-V$ index are from
L\"u et al. (1983) or, if not available, from L\"u (1991).  Column 6
gives a binarity flag (o: orbit, po: preliminary orbit with fixed
parameters, mp: minimum period only available, sb: suspected binary,
c: no radial-velocity variation). The last column refers to a note at
the end of the table providing the reference where the orbital
elements are to be found }
\label{tab2}
\begin{tabular}{lcrcclc}
\hline
HD/DM & Sp. & \multicolumn{1}{c}{Ba} & $V$ & $B-V$ & \multicolumn{1}{c}{bin.} 
& \multicolumn{1}{c}{orbit} \\ 
\hline
18182  & K0 &$<1$& 8.95 & 1.01 & sb & 0 \\
22589  & G5 &$<1$& 9.00 & 0.72 & po & 1 \\
26886  & G8 &  1 & 7.97 & 0.97 & o  & 2 \\
27271  & G8 &  1 & 7.49 & 1.00 & o  & 2 \\
40430  & K0 &  1 & 8.07 & 1.01 & mp & 1 \\
49841  & G8 &  1 & 8.56 & 0.99 & o  & 2 \\
50843  & K1 &  1 & 8.13 & 1.06 & c  & 0 \\
51959  & K2 &  1 & 8.95 & 1.06 & mp & 1 \\
53199  & G8 &  2 & 9.08 & 0.95 & po & 2 \\
58121  & K0 &  1 & 7.93 & 1.14 & o  & 2 \\
59852  & G9 &  1 & 8.67 & 0.92 & po & 1 \\
65699$^b$  & G5 &  1 & 5.07 & 1.12 &    & 0 \\
91208  & K0 &  1 & 8.02 & 0.95 & o  & 1 \\
95193  & K0 &  1 & 8.28 & 0.99 & o  & 1 \\
95345  & K2 &  1 & 4.83 & 1.16 & c  & 0 \\
101079 & K1 &  1 & 8.43 & 0.96 & mp & 2 \\
104979 & K0 &  1 & 4.16 & 0.99 & mp & 2 \\
119185 & K0 &  1 & 8.91 & 1.00 & c  & 0 \\
130255 & K0 &  1 & 8.61 & 1.08 & c  & 0 \\
131670$^a$ & K1 &  1 & 8.00 & 1.20 & o  & 1 \\
134698 & K1 &  1 & 8.72 & 1.31 & mp & 1 \\
139195 & K1 &  1 & 5.26 & 0.95 & o  & 12 \\  
143899 & G8 &  1 & 8.30 & 1.08 & o  & 1 \\
180622 & K1 &  1 & 7.62 & 1.23 & o  & 2 \\
183915 & K0 &  2 & 7.28 & 1.34 & sb & 0 \\
196673$^a$&K2& 2 & 6.98 & 1.12 & o  & 1 \\
200063 & K3 &  1 & 7.31 & 1.60 & o  & 2 \\
206778$^b$ & K2 &$<1$& 2.38 & 1.52 &    & 0 \\
210946 & K1 &  1 & 8.07 & 1.08 & o  & 2 \\
216219 & G5 &  1 & 7.45 & 0.74 & o  & 2 \\
218356 & K2 &  2 & 4.47 & 1.35 & sb & 0 \\
223617$^a$&K2& 2 & 6.94 & 1.16 & o  & 2 \\
288174 & K0 &  1 & 8.99 & 1.24 & o  & 1 \\
$-01^\circ3022$
       & K1 &  1 & 9.25 & 1.09 & o  & 1 \\
$-10^\circ4311$
       & G0 &  1 & 9.89 & 0.63 & mp & 1 \\
$-14^\circ2678$
       & K0 &$<1$& 9.81 & 0.99 & o  & 1 \\
\hline
\end{tabular}\\
$a$) {\small DAO}+\coravel\ measurements \\
$b$) misclassified as mild Ba star \\
References to orbital elements: 0: Jorissen et al. (1998);
1:~this paper; 2:~Paper~II; 12:~Griffin (1991)
\end{flushleft}
\end{table}

Among these mild barium stars, 27 stars are definitely binaries (3
with fixed parameters, and 6 with only a lower limit available on the
orbital period), 3 are suspected binaries, 4 show no evidence of
binary motion and 2 (\hd~65699 and \hd~206778) are supergiants
misclassified as mild barium stars (Smith \& Lambert 1987; superscript
'$b$' in Table~\ref{tab2}).  Results for 14 mild barium stars are
presented in Sect.~\ref{sect4}. The remaining 13 binaries are to be
found in Paper~II (12) and in Griffin (1991; 1).

\subsection{DAO stars}

For several barium stars monitored by McClure, a few recent \coravel\
measurements, obtained in the framework of other projects, are
nevertheless available. These new measurements often allow us to
improve the DAO orbit, since they significantly increase the span of
the monitoring.  All these new orbital parameters are used in the
analysis paper (Jorissen et al. 1998). The present paper and Paper~II
provide only four of these stars, 3 for which the number of \coravel\
measurements is fairly large (\hd~46407, \hd~131670 and \hd~223617)
and 1 (\hd~196673) for which the 2 new available observations
significantly change the period obtained by McClure \& Woodsworth
(1990). These stars are also included in Tables~\ref{tab1} and
\ref{tab2}, where they are identified by superscript '$a$' in
column~1. Their orbits are presented in Sect.~\ref{sect4} or in
Paper~II.



\subsection{Two barium stars of special importance}
\label{sect24}

The K0\,Ba3 star \hd~46407
and K2\,Ba2 star \hd~218356 (=56 Peg)
were included in the \coravel\ samples (Tables~1 and 2) because of
their unique photometric behaviour among barium stars. \hd~46407
exhibits long-term photometric variations in phase with the orbital
motion (Jorissen et al. 1991; Jorissen 1997), whereas 56 Peg is a
strong X-ray and UV source, indicating that it is an interacting
binary system possibly hosting an accretion disk (Schindler et
al. 1982; Dominy \& Lambert 1983).  A combined {\small DAO}/\coravel\
orbit of \hd~46407 is presented in Paper~II whereas no clear orbital
solution emerges for 56 Peg yet.

\subsection{Non-variable S stars}
\label{sect25}

An initial sample of 9 S stars, whose monitoring started in 1984 (with
preliminary results presented in Jorissen \& Mayor 1988, 1992) was
extended to 36 stars in 1988.  This sample contains bright northern S
stars from the {\it General Catalogue of Galactic S Stars} (GCGSS;
Stephenson 1984) with no variable star designation, neither in the
{\it General Catalogue of Variable Stars} (Kholopov et al. 1985) nor
in the {\it New Catalogue of Suspected Variable Stars} (Kukarkin et
al. 1982).  The criterion of photometric stability has been adopted to
avoid the confusion introduced by the envelope pulsations masking the
radial-velocity variations due to orbital motion.  Such a selection
criterion clearly introduces a strong bias against intrinsically
bright S stars.

The sample of 36 photometrically non-variable S stars is presented in
Table~\ref{tab3}. Among them, 24 are binaries\footnote{including
\hd~121447\ already presented in the strong barium sample} (6 have
only a lower limit available on the orbital period), 10 show no
evidence for binary motion, and 2 (\hd~262427 and
\bd~$+22^{\circ}4385$) are likely misclassified (Jorissen et
al. 1998).  The new spectroscopic binaries (18) are described in
Sect.~\ref{sect4} (16) and in Carquillat et al. (1998; 2).

\begin{table}[th!]
\begin{flushleft}
\caption[]{The sample of photometrically non-variable S stars monitored
with \coravel. The spectral type is from the GCGSS.  The $V$ magnitude
and $B-V$ index for S stars are from various sources, as listed in
{\it The General Catalogue of Photometric Data} (GCPD; Mermilliod et
al. 1997).  In case no data are listed in the GCPD, the $V$ magnitude
from the GCGSS is listed.  Column 6 gives a binarity flag (o:~orbit,
po:~preliminary orbit with fixed parameters, mp:~minimum period only
available, j:~jitter, c:~no radial-velocity variation). The last
column refers to a note at the end of the table providing the
reference where the orbital elements are to be found }
\label{tab3} 
\begin{tabular}{lr@{}llcclc}
\hline
\multicolumn{1}{c}{HD/DM}  & \multicolumn{2}{c}{{\small GCGSS}} 
& \multicolumn{1}{c}{Sp.\,\,\,\,} & $V$ & $B-V$ & \multicolumn{1}{c}{bin.} 
& \multicolumn{1}{c}{orbit} \\ 
\hline
7351   &   26 & & S3/2 & 6.43 & 1.70 & o  & 3 \\
30959  &  114 & & S3/1 & 4.74 & 1.84 & mp & 1 \\
35155  &  133 & & S4,1 & 6.77 & 1.80 & o  & 0,10 \\
49368  &  260 & & S3/2 & 7.65 & 1.78 & o  & 1 \\ 
61913  &  382 & & M3S  & 5.56 & 1.64 & j  & 0 \\
63733  &  411 & & S4/3 & 7.94 & 1.74 & o  & 1 \\ 
95875  &  720 & & S3,3 & 8.74 & 1.11 & o  & 1 \\
121447$^c$ &--& & S0   & 7.85 & 1.75 & o  & 1 \\
170970 & 1053 & & S3/1 & 7.60 & 2.00 & o  & 1 \\
184185 & 1140 & & S3*4 & 9.20 & 2.00 & mp & 1 \\ 
189581 & 1178 & & S4*2 & 8.70 & 2.00 & j  & 0 \\
191226 & 1192 & & M1-3S& 7.34 & 1.32 & o  & 3  \\ 
191589 & 1194 & & S    & 7.70 & 2.00 & o  & 1  \\
192446 & 1198 & & S6/1 & 9.80 & 2.00 & j  & 0  \\
216672 & 1315 & & S4/1 & 6.36 & 1.80 & j  & 0  \\ 
218634 & 1322 & & M4S  & 5.03 & 1.47 & mp & 1  \\ 
246818 &  156 & & S    & 9.60 & 2.00 & o  & 1 \\ 
262427$^b$ &  247 & &  & 9.90 & 2.00 &    & 0 \\
288833 &  233 & & S3/2 & 9.40 & 1.91 & mp & 1 \\ 
332077 & 1201 & & S3,1 & 9.00 & 1.96 & o  & 10 \\
343486 & 1092 & & S6,3 & --   & 2.00 & o  & 1 \\
$+04^\circ4354$
      & 1193 & & S4*3 &  9.50 & 2.00 & j  & 0 \\
$+15^\circ1200$
      &  219 & & S4/2 &  9.39 & 2.01 & j  & 0 \\ 
$+20^\circ4267$
      & 1158 & & Swk  & 10.80 & 2.55 & j  & 0 \\ 
$+21^\circ\ 255$
      &   45 & & S3/1 &  8.60 & 1.30 & o  & 1 \\
$+22^\circ\ 700$
      &   96 & & S6,1 & 10.50 & 2.00 & o  & 10 \\
$+22^\circ4385^b$
      & 1271 & &      & 10.10 & 2.00 &    & 0 \\
$+23^\circ3093$
      &  981 & & S3/3 &  9.90 & 2.00 & o  & 10 \\
$+23^\circ3992$
      & 1209 & & S3,3 & 10.50 & 2.00 & o  & 1 \\
$+24^\circ\ 620$
      &   87 & & S3/3 &  8.88 & 2.06 & o  & 10 \\
$+28^\circ4592$
      & 1334 & & S2/3:&  9.50 & 1.40 & o  & 1 \\
$+31^\circ4391$
      & 1267 & & S2/4 &  9.30 & 2.00 & mp & 1 \\
$+79^\circ\ 156$
      &  106 & & S4/2 & 10.10 & 1.70 & mp & 1 \\
$-04^\circ2121$
      &  416 & & S5/2 &  9.20 & 2.00 & j  & 0 \\
$-10^\circ1334$
      &  176 & & Sr   &  9.00 & 2.00 & j  & 0 \\
$-21^\circ2601$
      &  554 & & S3*3 &  9.20 & 2.00 & j  & 0 \\
\hline
\end{tabular}
$b$) star misclassified as S?\\
$c$) Two classifications: Ba5 or S0. Appears in both tables\\ 
References to orbital elements: 0:~Jorissen et al. (1998); 1:~this
paper; 3:~Carquillat et al. (1998); 10:~Jorissen \& Mayor (1992)
\end{flushleft}
\end{table}

\begin{table}[th!]
\begin{flushleft}
\caption[]{The sample of Mira S stars, SC stars and C stars with no
Tc lines, monitored with \coravel. The spectral type is from the GCGSS,
and $B-V$ from the GCPD }
\label{tab4}
\begin{tabular}{rr@{\hspace{1.mm}}lllc}
\hline
\multicolumn{1}{c}{HD/DM}  & \multicolumn{2}{c}{GCGSS} 
& \multicolumn{1}{c}{Var \,\,\,\,} & \,\,\,\,Sp. & $B-V$  \\ 
\hline
1967   &    9 & & R And      & S5-7/4-5e&1.97 \\
4350   &   12 & & U Cas      & S5/3e   & 2.00 \\
14028  &   49 & & W And      & S7/1e   & 1.70 \\ 
29147  &  103 & & T Cam      & S6/5e   & 2.30 \\ 
53791  &  307 & & R Gem      & S5/5    & 2.10 \\
70276  &  494 & & V Cnc      & S3/6e   & 2.20 \\ 
110813 &  803 & & S UMa      & S3/6e   & 2.10 \\ 
117287 & --\, & & R Hya      & M6e-M9eS& 1.60 \\ 
185456 & 1150 & & R Cyg      & S6/6e   & 2.00 \\
187796 & 1165 & & $\chi$ Cyg & S7/1.5e & 1.80 \\
190629 & 1188 & & AA Cyg     & S6/3    & 1.80 \\ 
195763 & 1226 & & Z Del      & S4/2e   & 2.00 \\
211610 & 1292 & & X Aqr      & S6,3e:  & 2.00 \\
\hline
286340 &  117 & & GP Ori     & SC7/8   & 2.90 \\ 
44544  &  212 & & FU Mon     & S7/7    & 3.00 \\ 
$-04^\circ1617$
       &  244 & & V372 Mon   & SC7/7   & 2.50 \\
$-08^\circ1900$
       &  344 & &            & S4/6    & 1.90 \\
54300  &--\,  & & R CMi  & CS   & 2.50 \\ 
198164 &--\,  & & CY Cyg & CS   & 2.00 \\ 
209890 &--\,  & & RZ Peg & CS   & 3.90 \\ 
\hline
46687  &--\,  & & UU Aur & C no Tc & 3.00 \\
76221  &--\,  & & X Cnc  & C no Tc & 3.36 \\
108105 &--\,  & & SS Vir & C no Tc & 3.00 \\
\hline
\end{tabular}
\end{flushleft}
\end{table}

\subsection{Mira S stars, SC stars and Tc-poor carbon stars}

A sample of 13 Mira S stars (Table~\ref{tab4}) has also been monitored
in order not to restrict the search for binaries to low-luminosity S
stars (see Sect.~\ref{sect25}). However, the envelope pulsations of
Mira stars seriously hamper that search by causing a substantial
radial-velocity jitter. This will be discussed in Sect.~\ref{sect5}.

A sample of 7 SC and CS stars suffering the same problem has been
monitored as well with \coravel, along with the 3 carbon stars lacking
Tc from the list of Little et al. (1987). These stars are also listed
in Table~\ref{tab4}. Radial-velocity variations are observed for all
these stars.

In case the photometric and radial-velocity periods are different, a
{\sl tentative} orbital solution is proposed. The results are
presented in Sect.~\ref{sect5}.

\begin{figure}
\ifigx{8.8 cm}{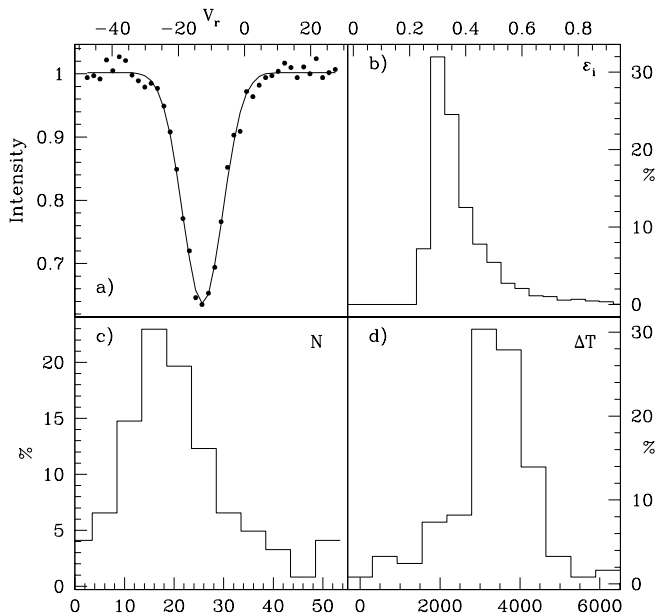}
\caption[]{Statistical overview of the combined samples of barium and
S stars: a)~typical cross-correlation profile (here an observation of
\hd~24035); b)~distribution of individual measurement errors
$\varepsilon_i$ (in {\kms}); c)~distribution of the number of
\coravel\ measurements per star $N$; d)~distribution of spans in days
($\Delta T$) between the first and the last measurements, for every
star}
\label{fig1}
\end{figure}

\section{Observations and overview of the survey}
\label{sect3}

Since 1983, more than 2500 measurements have been gathered for the
stars of the samples described above. The radial velocities are
obtained by usual cross-correlation (cc) and gaussian-fitting
techniques (see Baranne et al. 1979 or Duquennoy et al. 1991 for
details). An illustrative example of a profile obtained for \hd~24035
(cc-dip with the fitted gaussian) is displayed in Fig.~\ref{fig1}a.

The still unpublished individual measurements will be available at the
{\it Centre de Donn\'ees Stellaires} (CDS) in Strasbourg or on our
dedicated web page
(obswww.unige.ch/$\sim$udry/cine/barium/barium.html) and thus will not
appear here.

The mean precision of the measurements is about 0.3\,\kms\ (more than
76\% of the measurements between 0.25 and 0.4\,\kms) as shown in
Fig.~\ref{fig1}b displaying the distribution of individual errors
$\varepsilon_i$.  S stars with non-orbital radial-velocity variations
usually present non-gaussian cc-dips (see Sect.~\ref{sect:intrinsic})
leading to non-realistic error estimates that slightly swell the upper
tail of the distribution. Stars with broad cc-dip widths act in the
same way.

The distributions of the number of observations and of the time span
of the observations for every star are given in Figs.~\ref{fig1}c and
\ref{fig1}d, respectively. They illustrate the large observational
effort devoted to these programmes. The median value for the number of
observations per star is $\overline{N}=19$ which permits a good
estimate of the orbital parameters. The typical span
($\overline{\Delta T} =3398$\,d) is about 10 years ensuring a good
completeness of the orbital periods up to this value.

Further information on the \coravel\ observation and reduction
techniques can be found in Duquennoy et al. (1991).

\section{Radial-velocity curves and orbital parameters}
\label{sect4}

The radial-velocity monitoring of a binary star may lead to different
qualitative results depending on the parameters of the orbital
solution (period~$P$, eccentricity~$e$, amplitude~$K$), the number of
measurements or the observation sampling. The strategy for the
presentation of the results will thus depend on these different cases.
Situations possibly encountered are the following:

1) A stable solution is found and the orbital parameters can be
derived (flag `o' in Tables~\ref{tab2} and \ref{tab3}).  This is the
case for most of the orbits presented in this paper. A table is
provided with the orbital parameters and their uncertainties and a few
additional interesting related quantities as the number of
measurements ($N$) used to derive the orbital solution or the residue
($O-C$) around this solution. The phase-folded orbital solution is
then also displayed. Badly constrained parameters are readily
identifiable by their large uncertainties.

2) The orbital solution exists but is not fully constrained. One or
several parameters (usually $P$ or $e$) have to be fixed (e.g. for an
uncompletely-covered orbit or when the periastron passage in an
eccentric orbit has been missed; flag `po' in Tables~\ref{tab2} and
\ref{tab3}). In such a case, the adequate orbital parameter is fixed
to a probable value as given by the orbital solution with minimum
residuals.  The obtained {\sl minimized} orbital elements are given in
the same table as for case 1) but with no uncertainties on the fixed
parameter(s). A diagram with the velocity measurements folded in phase
is also provided.

3) The star has a clearly variable radial velocity but the period is
insufficiently covered (usually just a drift is observed) to derive a
preliminary orbit, even with fixed parameters (flag `mp' in
Tables~\ref{tab2} and \ref{tab3}). No solution is found. The minimum
period ($P>P_{\rm min}$) is indicated in the table of orbital
elements, along with the number of measurements and the time span of
the observations. The figure only displays radial velocities as a
function of Julian date. In addition to the individual measurements
available at the CDS (Sect.~\ref{sect3}), the analysis paper (Jorissen
et al. 1998) summarizes the interesting averaged quantities (radial
velocity and corresponding uncertainty, etc.) for these stars.

4) No orbital solution can be found because of other sources of
radial-velocity variations (jitter, flag `j' in Table~\ref{tab3})
masking a possible orbital motion.  Average quantities are given in
Jorissen et al (1998).
 
In the following subsections the results for the various star samples
described in Sect.\ref{sect2} will be presented and discussed in turn.

\begin{figure*}[th!]
\ifigx{17.25cm}{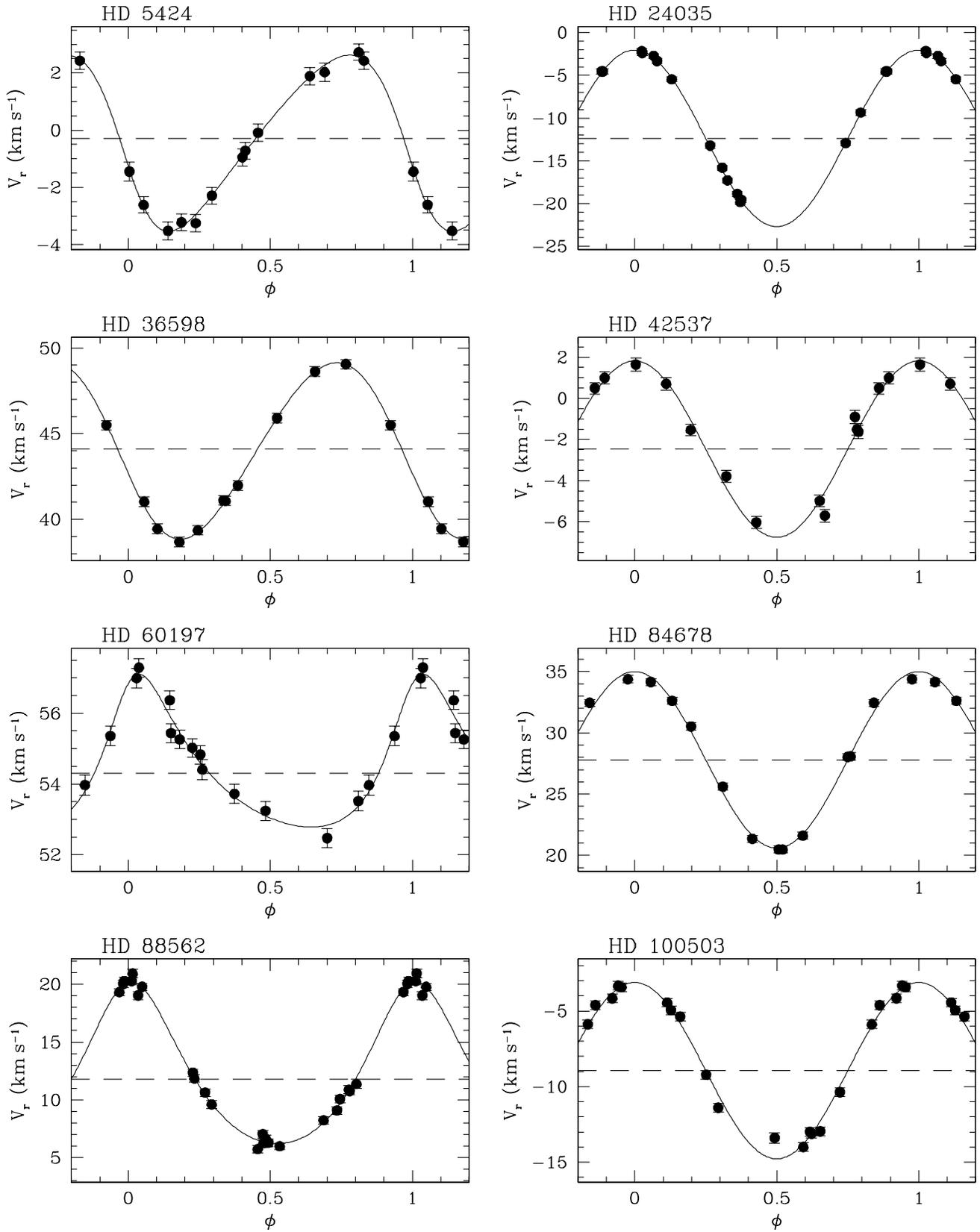}
\caption{Phase-folded radial-velocity curves for the strong barium
stars. {\sl Minimized} periods (see item 2 in Sect.~\ref{sect4}) were
fixed for \hd~123949 and \hd~211954 because of the non-complete
coverage of the orbits. \bd$+38^\circ$118 is a triple hierarchical system}
\label{fig2}
\end{figure*}

\addtocounter{figure}{-1}
\begin{figure*}[th!]
\ifigx{17.25cm}{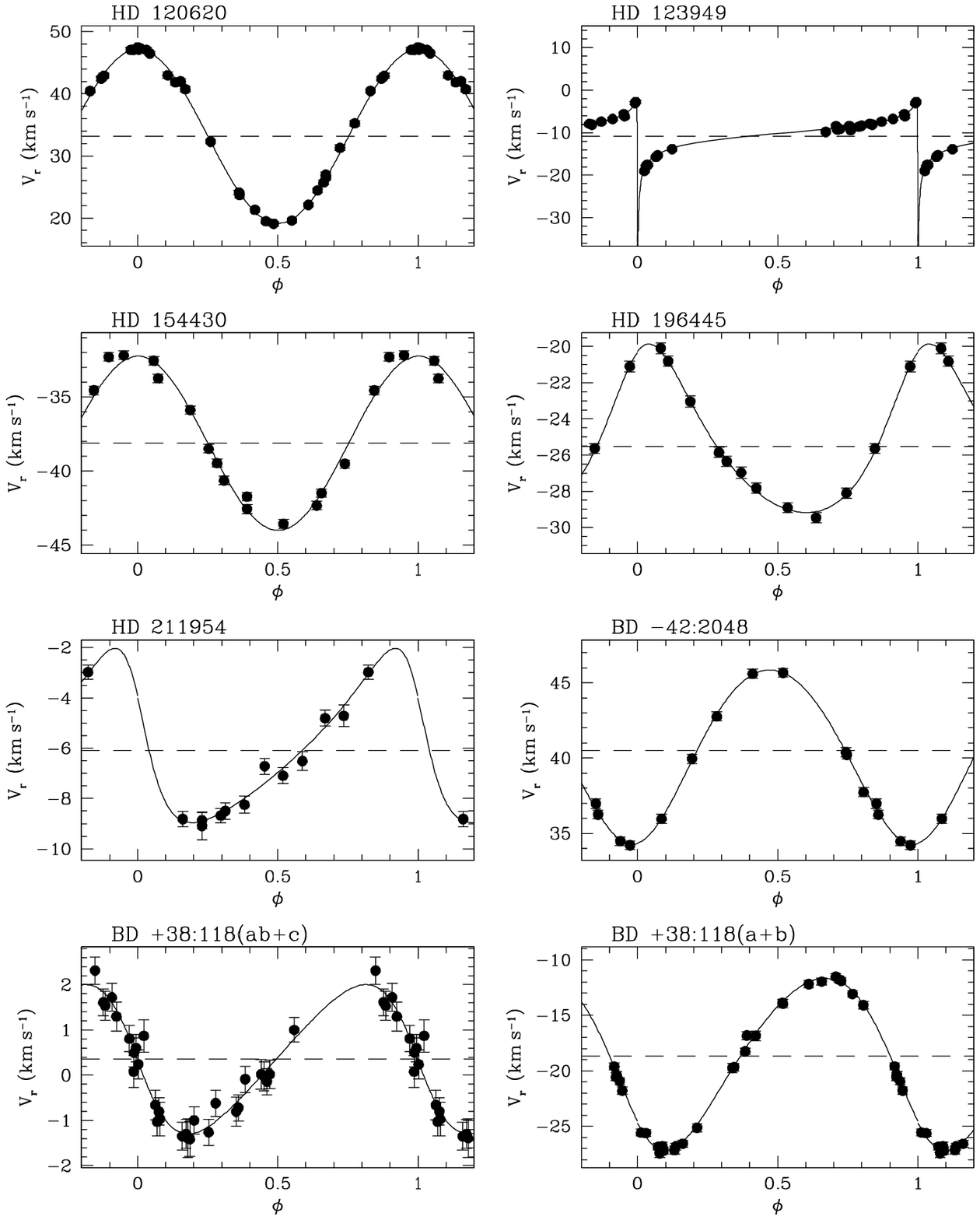}
\caption{(continued)}
\end{figure*}

\addtocounter{figure}{-1}
\begin{figure*}[th!]
\ifigx{17.25cm}{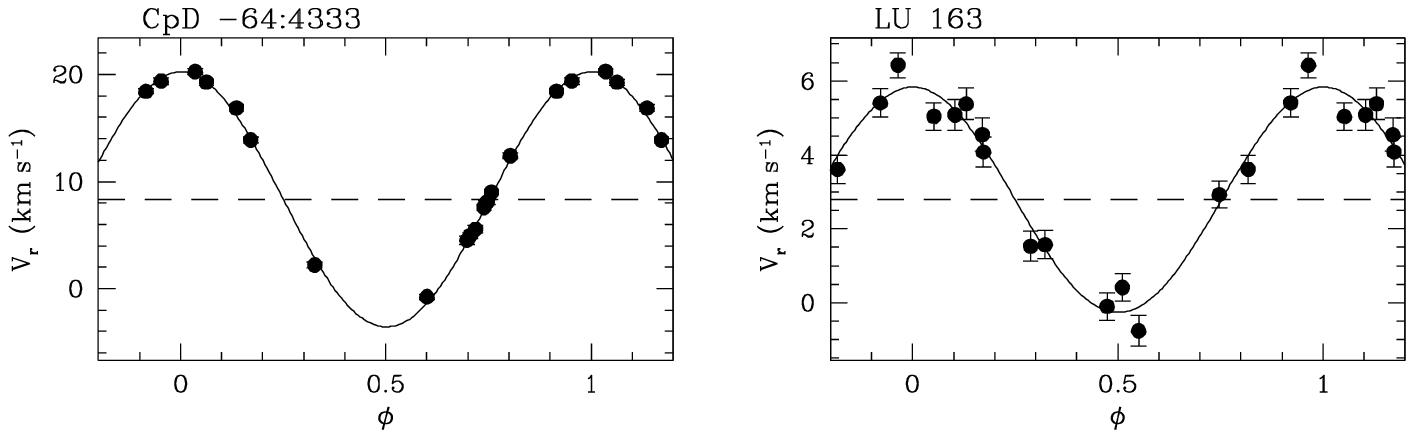}
\caption{(continued)}
\end{figure*}

\begin{table*}[th!]
\caption{Orbital elements for strong barium stars. No uncertainties
are given for fixed parameters. The symbol $>$ is used for
uncertainties exceeding the parameter values in case of badly
constrained orbits. $N$ is the number of measurements used to derive
the orbital solution and $O-C$ the residue around this
solution. $\Delta T$ is the span of the observations}
\label{tab5}
\begin{tabular}{|l|r|r|c|r@{\hspace{2mm}}l|r|r@{}r|r|r|c|c|c|}
\hline
\multicolumn{1}{|c|}{Id} & \multicolumn{1}{c|}{$P$} & \multicolumn{1}{c|}{$T$ [HJD} 
& \multicolumn{1}{c|}{$e$} & \multicolumn{2}{c|}{ $\gamma$} & \multicolumn{1}{c|}{$\omega$} 
& \multicolumn{2}{c|}{$K$} & \multicolumn{1}{c|}{$f(m)$} & \multicolumn{1}{c|}{ $a\sin{}i$} 
& \multicolumn{1}{c|}{$N$}  & \multicolumn{1}{c|}{$O-C$}  & \multicolumn{1}{c|}{$\Delta T$} \\
\multicolumn{1}{|c|}{HD/DM/..} & \multicolumn{1}{c|}{[days]} & \multicolumn{1}{c|}{$-2400000$]} 
& \multicolumn{1}{c|}{} 
& \multicolumn{2}{c|}{[\kms]} & \multicolumn{1}{c|}{$[\deg]$} & \multicolumn{2}{c|}{[\kms]} 
& \multicolumn{1}{c|}{}  & \multicolumn{1}{c|}{[Gm]} & \multicolumn{1}{c|}{ } 
& \multicolumn{1}{c|}{[\kms]} & \multicolumn{1}{c|}{[days]} \\
\hline
5424 &1881.53 &46202.80 &0.226 &$-0.28$ &  &104.63 &3.08 & &5.281e-03 &77.64 &13 &0.180 &3306 \\
     &  18.59 &   34.86 &0.036 & 0.05   &  &  5.87 &0.08 & &4.221e-04 & 2.19 & & & \\
\hline
24035 &377.82 &48842.65 &0.020 &$-12.51$&  &214.82 &\hspace{2mm}10.64 & &4.725e-02 &55.27 &15 &0.186 &3271\\
      &  0.35 &   26.14 &0.010 &  0.13  &  & 24.76 & 0.27 & &3.544e-03 & 0.14 & & & \\
\hline
36598 &2652.81 &45838.95 &0.084 &44.10  &  &104.59 &5.15 & &3.715e-02 &187.06 &11 &0.210 &3270\\
      &  22.66 &  107.50 &0.018 & 0.07  &  & 13.96 &0.09 & &2.060e-03 &  3.77 & & & \\
\hline
42537 &3216.22 &46147.32 &0.156 &$-2.47$ & &237.61 &4.40 & &2.741e-02 &192.18 &12 &0.432 &3270\\
      &  54.68 &  162.50 &0.047 & 0.16   & & 19.27 &0.25 & &4.722e-03 & 11.46 & & & \\
\hline
60197 &3243.76 &46015.97 &0.340 &54.30   & &330.94 &2.16 & &2.829e-03 &90.66 &14 &0.309 &3970\\
      &  66.34 &   98.30 &0.051 & 0.10   & & 10.89 &0.13 & &5.542e-04 & 6.17 & & & \\
\hline
84678 &1629.91 &44512.25 &0.062 &27.92   & &190.00 &7.16 & &6.167e-02 &160.08 &12 &0.297 &3270\\
      &  10.38 &   81.44 &0.020 & 0.11   & & 17.94 &0.12 & &3.157e-03 &  2.89 & & & \\
\hline
88562 &1445.05 &45781.71 &0.204 &11.79   & &353.43 &6.98 & &4.787e-02 &135.77 &23 &0.442 &3253\\
      &   8.53 &   35.50 &0.017 & 0.10   & &  7.84 &0.12 & &2.590e-03 &  2.56 & & & \\
\hline
100503 &554.41 &46144.83 &0.061 &$-8.9$  & &358.06 &5.80 & &1.116e-02 &44.12 &16 &0.549 &3271\\
       &  1.91 &   54.82 &0.045 &0.15    & & 34.83 &0.24 & &1.366e-03 & 1.81 & & & \\
\hline
120620 &217.18 &48831.22 &0.010 &33.21   & &172.87 &14.06 & &6.226e-02 &41.98 &28 &0.421 &3253\\
       &  0.08 &   27.42 &0.009 & 0.09   & & 45.50 & 0.11 & &1.485e-03 & 0.33 & & & \\
\hline
123949 &9200.0 &49144.96 &0.972 &$-10.82$& &128.51 &20.45 & &1.055e-01 &606.86 &25 &0.301 &4396\\
       &\multicolumn{1}{c|}{--}    &   62.89 &0.057 &  0.26  & & 49.01 &\multicolumn{2}{c|}{$>$}  &\multicolumn{1}{c|}{$>$}  &122.42 & & & \\
\hline
154430 &1668.11 &47442.02 &0.108 &$-38.14$ & &312.00 &5.85 & &3.400e-02 &133.30 &15 &0.480 &3058\\
       &  17.36 &   76.79 &0.031 &  0.14   & & 16.55 &0.19 & &3.315e-03 &  4.53 & & & \\
\hline
196445 &3221.35 &46037.95 &0.237 &$-25.53$ & &335.85 &4.66 & &3.101e-02 &200.46 &12 &0.226 &3306\\
       &  43.00 &   64.62 &0.023 &  0.09   & &  5.31 &0.11 & &2.226e-03 &  5.42 & & & \\
\hline
211954 &5000.0 &45497.96 &0.391 &$-6.08$ & &64.32 &3.47 & &1.686e-02 &219.33 &14 &0.353 &4081\\
       &\multicolumn{1}{c|}{--}  &149.37 &0.080 & 0.13   & &10.07 &0.31 & &4.934e-03 & 21.40 & & & \\
\hline
$+38^\circ \ 118$       &299.37 &47624.61 &0.144 &$-18.68$ & &132.44 &7.76 & &1.406e-02 &31.60 &30 &0.303 &4260\\
(a+b) &  0.19 &    3.15 &0.013 &  0.06   & &  3.93 &0.08 & &4.205e-04 & 0.32 & & & \\
\hline
$+38^\circ \ 118$ &3876.66 &46757.73 &0.208 &$-18.32$ & &91.17 &1.65 & &1.704e-03 &86.23 &30 &0.293 &4260\\
(ab+c)& 112.24 &  136.54 &0.062 &0.09 & &14.56 &0.11 & &3.616e-04 & 6.54 & & & \\
\hline
$-42^\circ$2048 &3259.96 &46948.95 &0.080 &40.52 & &188.85 &5.79 & &6.496e-02 &258.56 &12 &0.239 &3270\\
                &  28.30 &  142.67 &0.016 & 0.08 & & 16.40 &0.11 & &3.711e-03 &  5.36 & & & \\
\hline
$-64^\circ$4333 &386.04 &47486.99 &0.031 &8.33 & &85.40 &11.96 & &6.840e-02 &63.43 &16 &0.314 &3309\\
CpD             &  0.48 &   31.72 &0.012 &0.19 & &29.51 & 0.20 & &3.426e-03 & 1.06 & & & \\
\hline
L\"u$\;$163 &965.15 &47283.53 &0.035 &2.78 & &351.32 &3.06 & &2.879e-03 &40.64 &14 &0.572 &4036\\
            & 16.01 &  350.26 &$>$ &0.16 & &132.49 &0.22 & &6.199e-04 & 2.99 & & & \\
\hline
\end{tabular}
\end{table*}

\begin{figure*}[th!]
\ifigx{17.25cm}{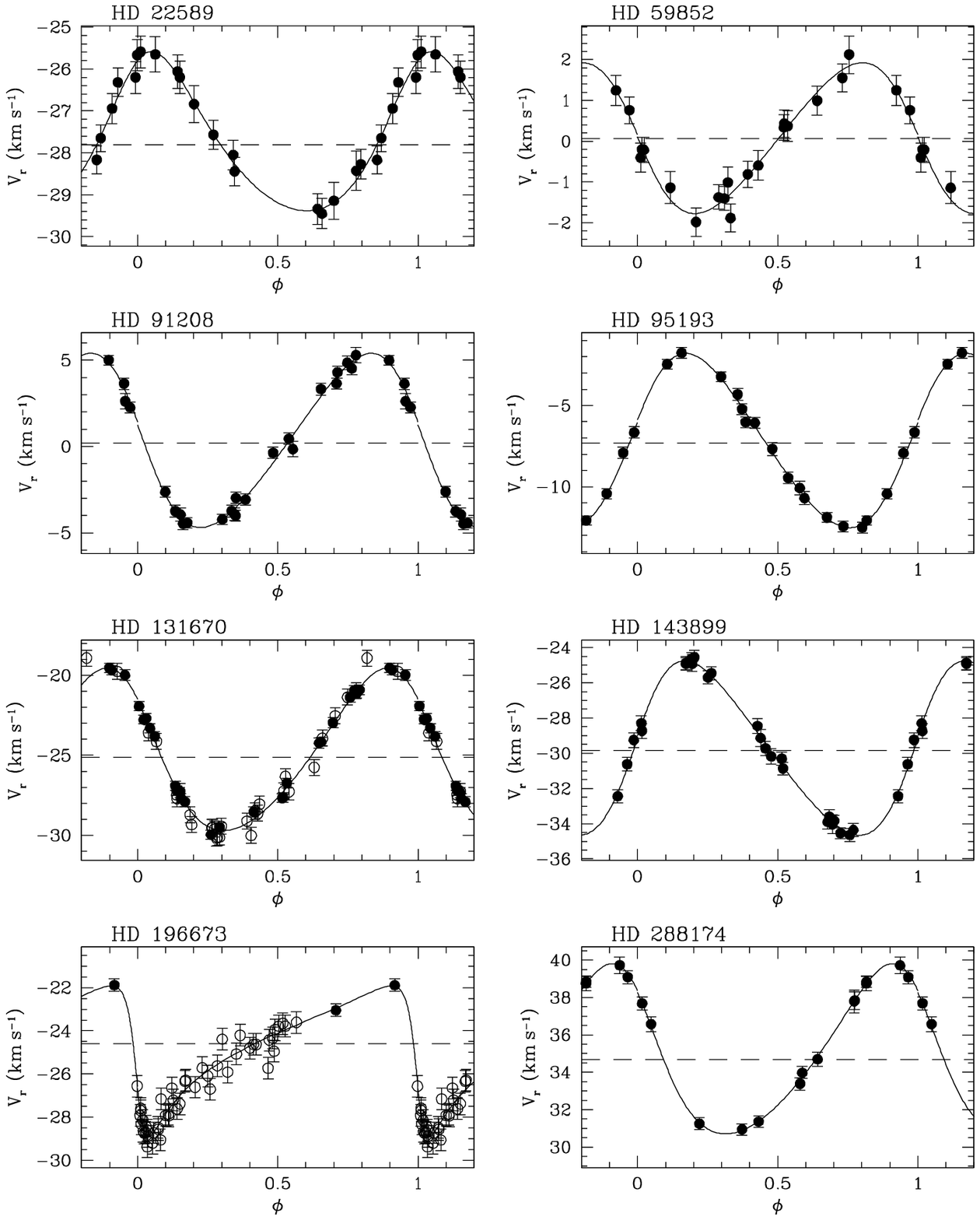}
\caption{Phase-folded radial-velocity curves for mild barium stars
with orbital solutions.  Open circles are for DAO measurements.  Two
new \coravel\ measurements allow us to propose a new minimum-period
estimate for \hd~196673.  Long-period stars without orbital solutions
have their radial velocities displayed as a function of Julian dates}
\label{fig3}
\end{figure*}

\addtocounter{figure}{-1}
\begin{figure*}[th!]
\ifigx{17.25cm}{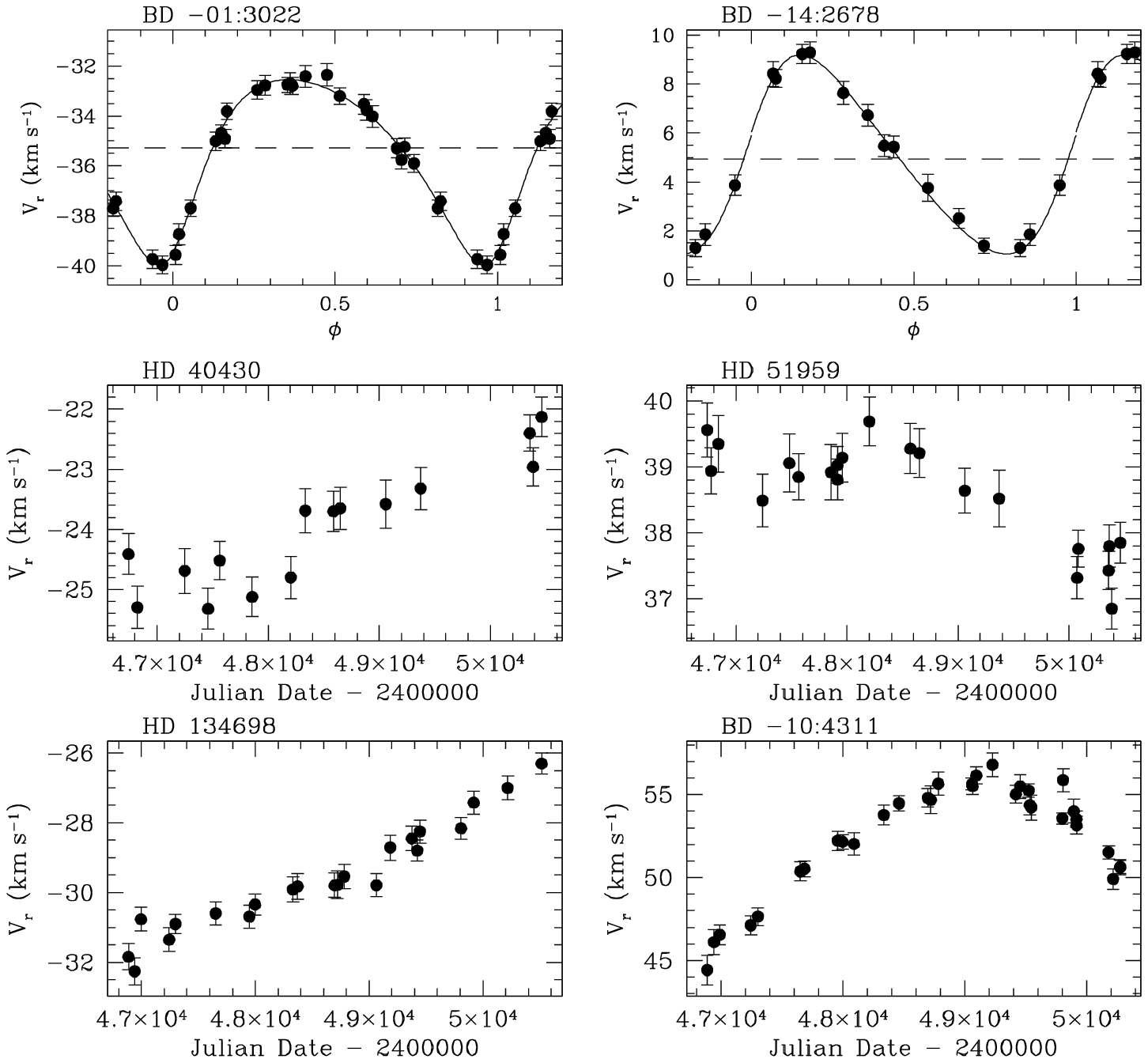}
\caption{(continued)}
\end{figure*}

\begin{figure*}[th!]
\ifigx{17.25cm}{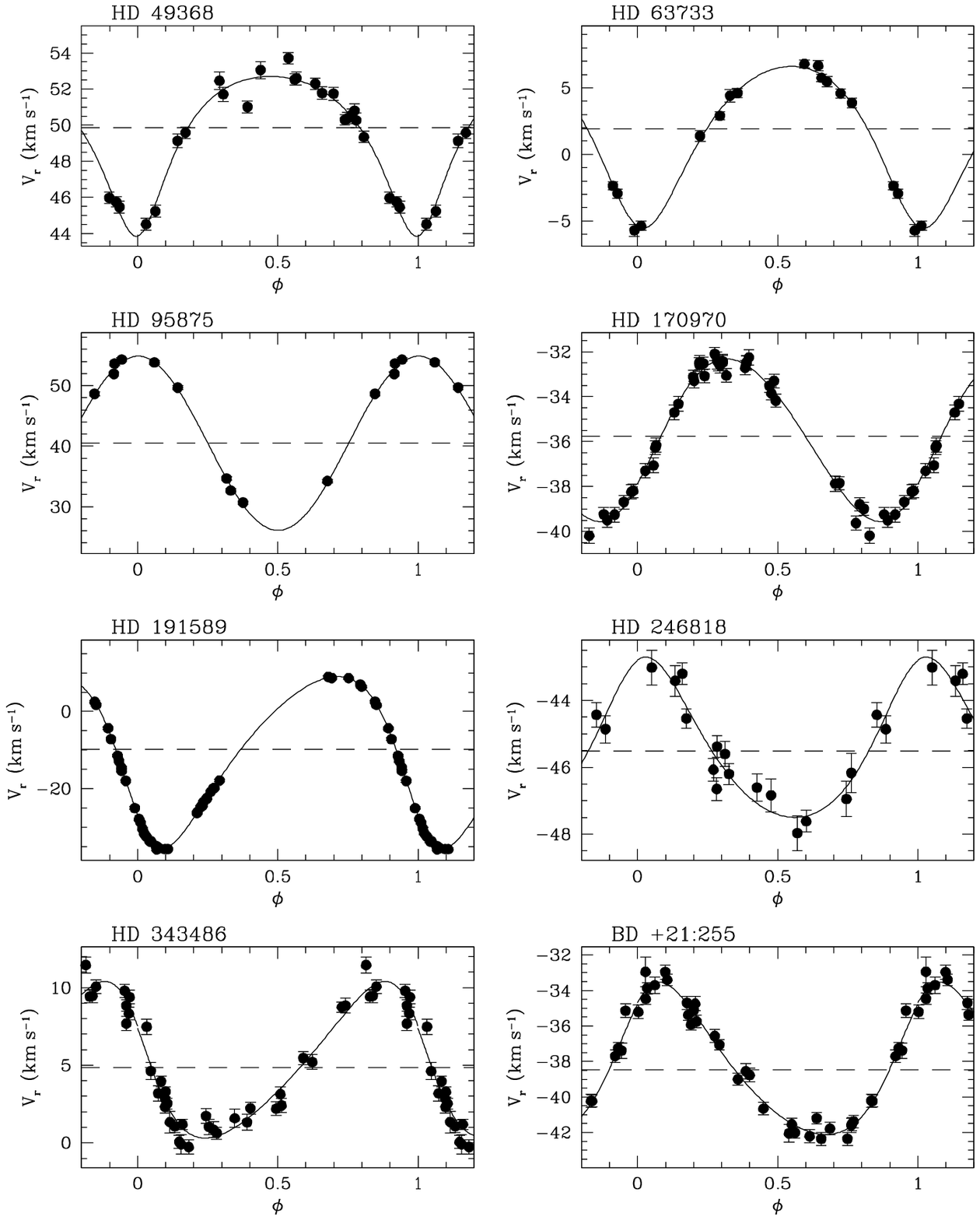}
\caption{Phase-folded radial-velocity curves of the photometrically
non-variable S stars with orbital solutions.  Long-period stars
without orbital solutions have their radial velocities displayed as a
function of Julian dates}
\label{fig4}
\end{figure*}

\addtocounter{figure}{-1}
\begin{figure*}[th!]
\ifigx{17.25cm}{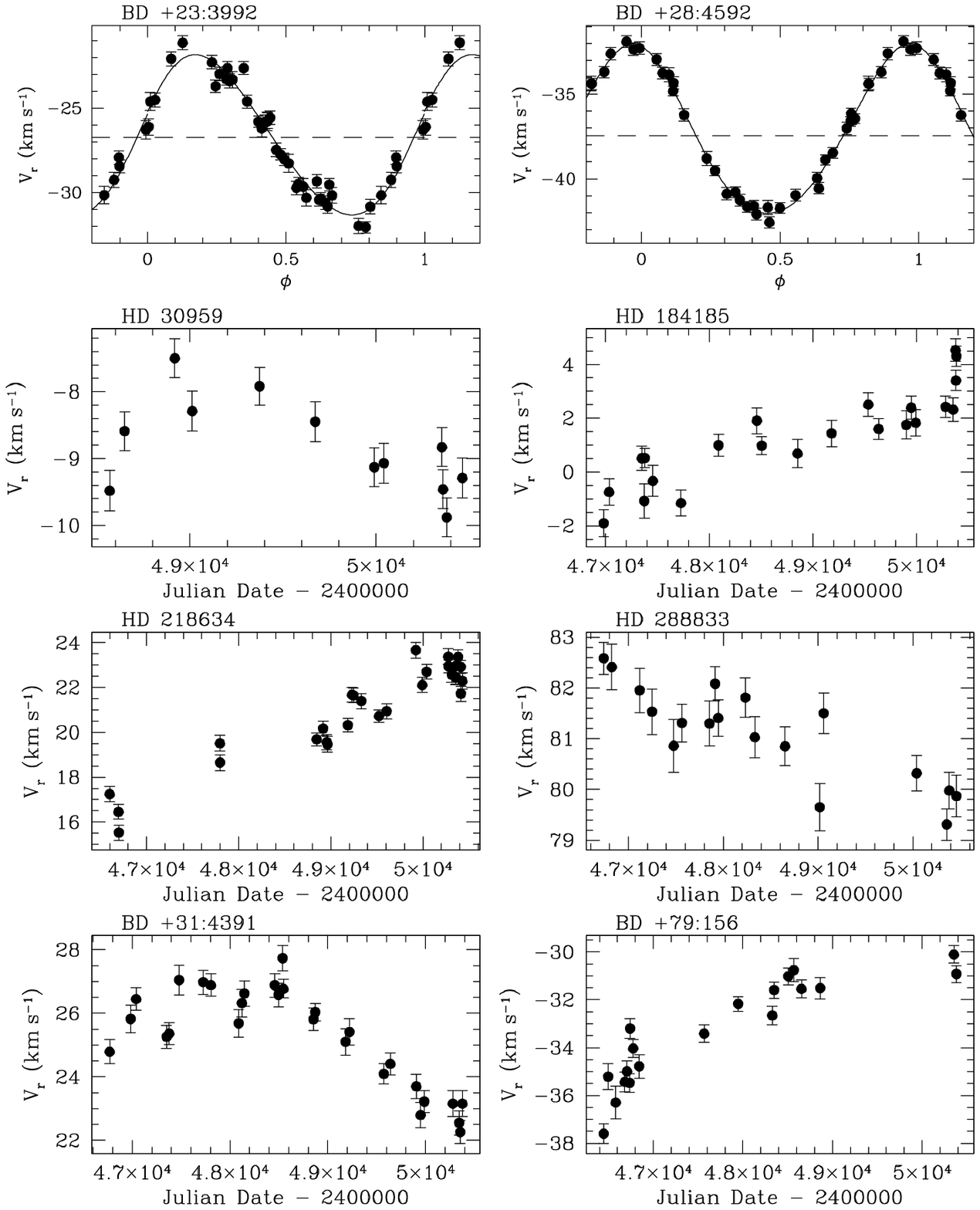}
\caption{(continued)}
\end{figure*}

\begin{table*}[th!]
\caption{Orbital elements for mild barium stars. No uncertainties
are given for fixed parameters. $N$ is the number of measurements used
to derive the orbital solution and $O-C$ the residue around this
solution. $\Delta T$ is the span of the observations }
\label{tab6}
\begin{tabular}{|l|r|r|l|r@{\hspace{2mm}}l|r|c|r|r|c|c|c|}
\hline
\multicolumn{1}{|c|}{Id} & \multicolumn{1}{c|}{$P$} & \multicolumn{1}{c|}{$T$ [HJD} 
& \multicolumn{1}{c|}{$e$} & \multicolumn{2}{c|}{$\gamma$} & \multicolumn{1}{c|}{$\omega$} 
& \multicolumn{1}{c|}{$K$} & \multicolumn{1}{c|}{$f(m)$} & \multicolumn{1}{c|}{ $a\sin{}i$} 
& \multicolumn{1}{c|}{ N}  & \multicolumn{1}{c|}{$O-C$}  & \multicolumn{1}{c|}{$\Delta T$}\\
\multicolumn{1}{|c|}{HD/DM} & \multicolumn{1}{c|}{[days]} & \multicolumn{1}{c|}{$-2400000$]} 
& \multicolumn{1}{c|}{} 
& \multicolumn{2}{c|}{[\kms]} & \multicolumn{1}{c|}{$[\deg]$} & \multicolumn{1}{c|}{[\kms]} 
& \multicolumn{1}{c|}{}  & \multicolumn{1}{c|}{[Gm]} & \multicolumn{1}{c|}{ } 
& \multicolumn{1}{c|}{[\kms]} & \multicolumn{1}{c|}{[days]}\\
\hline
22589 & 5721.181 &48674.73 &0.240 &$-27.97$ & &341.09 &1.97 &4.180e-03 &150.74 &19 & 0.217 &3641\\
      &  453.986 &  133.44 &0.175 &  0.62   & & 17.38 &0.28 &2.471e-03 & 63.26 & &  & \\
\hline
40430 &$P>3700$  & & & & & & & & &15 & &3720 \\
\hline
51959 &$P>3700$  & & & & & & & & &21 & &3720 \\
\hline
59852 &3463.906 &46841.03 &0.152 &0.07 & &88.83 &1.85 &2.190e-03 &86.98 &19 & 0.273 &3643\\
      &  53.776 &  187.14 &0.057 &0.07 & &20.68 &0.12 &4.376e-04 & 5.93 & & & \\
\hline
91208 &1754.002 &45628.36 &0.171 &0.19 & &79.25 &5.05 &2.239e-02 &119.93 &24 & 0.387 &3036\\
      &  13.321 &   54.16 &0.022 &0.08 & &10.11 &0.11 &1.555e-03 &  2.91 & & & \\
\hline
95193 &1653.728 &46083.62 &0.135 &$-7.32$ & &283.23 &5.37 &2.585e-02 &120.97 &18 & 0.264 &2596\\
      &   8.950 &   38.25 &0.017 & 0.07   & &  7.60 &0.10 &1.446e-03 &  2.34 & & &\\
\hline
131670 &2929.730 &46405.11 &0.162 &$-25.13$ & &49.88 &5.16 &4.005e-02 &204.94 &55 & 0.363 &5865\\
       &  12.265 &   44.96 &0.014 &   0.05  & & 5.44 &0.08 &1.830e-03 &  3.22 & & & \\
\hline
134698 &$P>3600$ & & & & & & & & &22 & &3630 \\
\hline
143899 &1461.608 &46243.43 &0.194 &$-29.83$ & &277.66 &4.95 &1.740e-02 &97.64 &26 &0.271 &2649\\
       &   6.919 &   19.76 &0.025 &  0.06   & &  4.57 &0.10 &1.065e-03 & 2.04 & & & \\
\hline
196673 &6500.0 &43953.72 &0.645 &$-24.61$ & &111.06 &3.49 &1.274e-02 &237.99 &51 &0.468 &5971 \\
       &\multicolumn{1}{c|}{--}  &   26.79 &0.028 &  0.10   & &  4.48 &0.17 &2.175e-03 & 13.54 & & & \\
\hline
288174 &1824.286 &47157.62 &0.194 &34.67 & &48.33 &4.55 &1.684e-02 &111.96 &14 &0.147 &3296\\
       &   7.057 &   17.73 &0.015 & 0.04 & & 3.95 &0.07 &7.431e-04 &  1.70 & & & \\
\hline
$-01^\circ 3022$ &3252.530 &46817.14 &0.285 &$-35.38$ & &212.09 &3.76 &1.586e-02 &161.34 &26 & 0.253 &3667\\
                 &  31.420 &   34.35 &0.018 &  0.06   & &  4.31 &0.09 &1.185e-03 &  4.28 & & & \\
\hline
$-10^\circ 4311$ &$P>3400$  & & & & & & & & &33 & &3392 \\
\hline
$-14^\circ 2678$ &3470.519 &48828.06 &0.217 &4.93 & &282.61 &4.07 &2.252e-02 &189.38 &15 &0.390 &3295\\
                 & 107.690 &  107.22 &0.040 &0.12 & & 12.16 &0.14 &2.501e-03 &  8.94 & & & \\
\hline
\end{tabular}
\end{table*}

\begin{table*}[th!]
\caption{Orbital elements for the S stars with no (strong) light variations.
The symbol $>$ is used for uncertainties exceeding the parameter
values in case of badly constrained orbits. $N$ is the number of
measurements used to derive the orbital solution and $O-C$ the residue
around this solution. $\Delta T$ is the span of the observations}
\label{tab7}
\begin{tabular}{|l|r|r|l|r@{\hspace{2mm}}l|r|r@{}r|r|r|c|c|c|}
\hline
\multicolumn{1}{|c|}{Id} & \multicolumn{1}{c|}{$P$} & \multicolumn{1}{c|}{$T$ [HJD} 
& \multicolumn{1}{c|}{$e$} & \multicolumn{2}{c|}{ $\gamma$} & \multicolumn{1}{c|}{$\omega$} 
& \multicolumn{2}{c|}{$K$} & \multicolumn{1}{c|}{$f(m)$} & \multicolumn{1}{c|}{ $a\sin{}i$} 
& \multicolumn{1}{c|}{ N}  & \multicolumn{1}{c|}{$O-C$}  & \multicolumn{1}{c|}{$\Delta T$}\\
\multicolumn{1}{|c|}{HD/DM} & \multicolumn{1}{c|}{[days]} & \multicolumn{1}{c|}{$-2400000$]} 
& \multicolumn{1}{c|}{} 
& \multicolumn{2}{c|}{[\kms]} & \multicolumn{1}{c|}{$[\deg]$} & \multicolumn{2}{c|}{[\kms]} 
& \multicolumn{1}{c|}{}  & \multicolumn{1}{c|}{[Gm]} & \multicolumn{1}{c|}{ } 
& \multicolumn{1}{c|}{[\kms]} & \multicolumn{1}{c|}{[days]}\\
\hline
30959 &$P>1900$ & & & & & & & & & &12 & &1895 \\
\hline
49368 &2995.903 &45145.37 &0.357 &49.85 & &184.37 &4.43 & &2.201e-02 &170.39 &23 &0.577 &3635\\
      &  67.152 &   65.01 &0.048 & 0.13 & &  7.27 &0.20 & &3.083e-03 &  8.73 & & &\\
\hline
63733 &1160.701 &45990.92 &0.231 &1.91 & &168.22 &6.08 & &2.492e-02 &94.38 &14 &0.379 &2685\\
      &   8.922 &   42.60 &0.034 &0.12 & &  8.91 &0.21 & &2.628e-03 & 3.39 & & &\\
\hline
95875 &197.236 &48843.35 &0.023 &40.61 & &25.86 &\hspace{2mm}14.21 & &5.878e-02 &38.54 &10 &0.699 &1826\\
      &  0.365 &   91.98 &\multicolumn{1}{c|}{$>$}   & 1.06 & &\multicolumn{1}{c|}{$>$}  & 2.12 & &9.405e-03 & 2.06 & & &\\
\hline
170970 &4391.997 &48213.21 &0.084 &$-35.77$ & &237.10 &3.62 & &2.133e-02 &217.58 &39 &0.332 &3463\\
       & 201.991 &  277.15 &0.043 &  0.06   & & 24.40 &0.09 & &1.856e-03 & 11.35 & & &\\
\hline
184185 &$P>3400$ & & & & & & & & & &22 & &3403 \\
\hline
191589 &377.342 &48844.02 &0.253 &$-9.74$ & &128.63 &22.30 & &3.937e-01 &111.96 &41 &0.292 &1895\\
       &  0.138 &    0.99 &0.003 & 0.08   & &  1.15 & 0.09 & &5.052e-03 &  0.48 & & &\\
\hline
218634 &$P>3700$ & & & & & & & & & &28 & &3780 \\
\hline
246818 &2548.543 &47115.84 &0.182 &$-45.52$ & &344.20 &2.40 & &3.473e-03 &82.66 &17 &0.595 &3718\\
       &  73.188 &  234.81 &0.111 &  0.19   & & 35.59 &0.34 & &1.494e-03 &12.06 & & &\\
\hline
288833 &$P>3900$ & & & & & & & & & &18 & &3721 \\
\hline
343486 &3165.723 &46880.50 &0.241 &4.86 & &65.93 &5.05 & &3.862e-02 &213.21 &37 &0.823 &3398\\
       &  37.626 &   89.02 &0.035 &0.15 & &10.78 &0.20 & &4.804e-03 &  9.16 & & &\\
\hline
$+21^\circ\ 255$ &4137.166 &43578.31 &0.209 &$-38.48$ & &316.59 &4.29 & &3.178e-02 &238.82 &36 &0.511 &4068\\
                 & 316.831 &  260.49 &0.043 &  0.27   & & 11.95 &0.16 & &4.321e-03 & 20.36 & & &\\
\hline
$+23^\circ 3992$ &3095.574 &45365.62 &0.105 &$-26.72$ & &286.42 &4.75 & &3.380e-02 &200.90 &43 &0.706 &4135\\
                 &  41.740 &  177.00 &0.034 &  0.12   & & 19.87 &0.19 & &4.011e-03 &  8.35 & & &\\
\hline
$+28^\circ 4592$ &1252.941 &48161.32 &0.091 &$-37.47$ & &13.80 &5.01 & &1.617e-02 &85.97 &34 &0.316 &3611\\
                 &   3.538 &   33.83 &0.017 &  0.06   & & 9.92 &0.08 & &8.147e-04 & 1.46 & & &\\
\hline
$+31^\circ 4391$ &$P>3600$ & & & & & & & & & &28 & &3607 \\
\hline
$+79^\circ\ 156$ &$P>3900$ & & & & & & & & & &19 & &3936 \\
\hline
\end{tabular}
\end{table*}

\begin{table*}[th!]
\caption{Orbital or {\sl pseudo-}orbital (pulsational) elements for
Mira S stars, SC stars and Tc-poor C stars.  The most likely cause of
the radial-velocity variations is given in column 2: `orb' for orbital
motion and `puls' for intrinsic atmospheric phenomenon. No
uncertainties are given for fixed parameters. The symbol $>$ is used
for uncertainties exceeding the parameter values in case of badly
constrained orbits. $N$ is the number of measurements used to derive
the orbital solution and $O-C$ the residue around this
solution. $\Delta T$ is the span of the observations}
\label{tab8}
\begin{tabular}{|l|c|c|r|r|c|r@{\hspace{2mm}}l|r|c|r|c|c|c|}
\hline
\multicolumn{1}{|c|}{HD/DM} &Var
& \multicolumn{1}{c|}{$P_{phot}$} 
& \multicolumn{1}{c|}{$P_{orb}$} & \multicolumn{1}{c|}{$T$ [HJD} 
& \multicolumn{1}{c|}{$e$} & \multicolumn{2}{c|}{ $\gamma$} & \multicolumn{1}{c|}{$\omega$} 
& \multicolumn{1}{c|}{$K$} & \multicolumn{1}{c|}{$f(m)$} 
& \multicolumn{1}{c|}{ N}  & \multicolumn{1}{c|}{$O-C$}  & \multicolumn{1}{c|}{$\Delta T$}\\
\multicolumn{1}{|c|}{GCVS} & & \multicolumn{1}{c|}{[days]} 
& \multicolumn{1}{c|}{[days]} 
& \multicolumn{1}{c|}{$-2400000$]} & \multicolumn{1}{c|}{} & \multicolumn{2}{c|}{[\kms]} 
& \multicolumn{1}{c|}{$[\deg]$} & \multicolumn{1}{c|}{[\kms]} & \multicolumn{1}{c|}{}  
& \multicolumn{1}{c|}{ } & \multicolumn{1}{c|}{[\kms]} & \multicolumn{1}{c|}{[days]}\\
\hline
76221  &orb? &$\sim 195$ &491.4	  &48642.17  &0.17  &$-5.96$ & &258.3  & 1.37  &1.270e-02 &15  &0.664  &2117 \\
X Cnc  &     &      &\multicolumn{1}{c|}{--}  	  &   87.15  &$>$   & 0.20   & &65.1 & 0.28  &7.748e-05 & & & \\
\hline
110813 &orb &225.9 &592.9 &49120.14 &0.00 &1.41 & &0.0 &6.26 &1.508e-02 &13 &1.126 &2667\\
S UMa  &    &     & 61.8 &   98.27 &\multicolumn{1}{c|}{--}  &3.33 & &\multicolumn{1}{c|}{--}  &4.25 &3.036e-03 & & &\\
\hline
$-08^\circ 1900$ &orb? &$\sim59$ &544.2 &48590.38 &0.55 &72.257 & &216.8 &3.24 &1.119e-02 &17 &0.732 &3619\\
              &   &   &  5.7 &   33.46 &\multicolumn{1}{c|}{--} & 0.33  & &15.3 & 0.85 &8.841e-04 & & &\\
\hline
54300  &puls &337.8 &337.3 &48693.12 &0.39 &44.03 & & &8.04 & &16 &1.836 &3740 \\
R CMi         &     &      &  1.3 &   10.20 &0.08 & 0.62 & & &0.77 & & & & \\
\hline
108105 &puls &$\sim 364$ & 361.2  &50199.59  &0.28  &4.06  & & &4.40 & &16 & 1.429  &1976 \\
SS Vir   &    &      &  3.4   & 35.94    &0.12  &0.73  & & &0.48 & & & & \\
\hline
209890 &puls &438.7 &437.3 &48413.34 &0.15 &$-32.28$ & & &8.96 & &21 &2.249 &2932 \\
RZ Peg   &     &      &  3.9 &   34.61 &0.10 &  0.52   & & &0.82 & & & & \\
\hline
\end{tabular}
\end{table*}

\subsection{Strong barium stars}

Definitive or preliminary orbits have been obtained for all strong
barium stars except \hd~19014. The orbital parameters are given in
Table~\ref{tab5}.  For two uncompletely-covered orbits (\hd~123949 and
\hd~211954), the periods have been fixed to {\sl minimized} values
(see item 2 above).  In 8 cases, even though the Lucy-Sweeney test
was compatible with a circular orbit at a 5\% confidence level (Lucy
\& Sweeney 1971), the slightly `eccentric' orbit has been listed,
because in the case of barium stars, there is no {\it physical}
argument to prefer the circular orbit (see Jorissen et al. 1998).
The corresponding radial-velocity measurements folded in phase are
displayed in Fig.~\ref{fig2}.

\bd~$+38^\circ$118 is a triple hierarchical system. The long-period
orbit (noted ab+c) describes the motion of the c component, relatively
to the center of mass of the close (short-period) system (noted
a+b). The orbits are obtained iteratively by correcting the
short-period orbital motion from the long-period perturbation.

\subsection{Mild barium stars}

Orbital elements were derived for 10 among the 14 mild barium stars
presented in this section (Table~\ref{tab6}), the others only allowing
minimum period estimates. The corresponding phase-folded curves are
displayed in Fig.~\ref{fig3}, along with the temporal radial-velocity
variations of the remaining 4 stars with no orbital solution.

The star \hd~196673 deserves a special note. It belongs to the DAO
sample with a spectroscopic orbit published by McClure \& Woodsworth
(1990). However, as shown by 2 new \coravel\ measurements, the
inferred orbital period ($P=4000$\,d) was too short.  Based on all the
{\small DAO}+\coravel\ measurements, we propose a new period (fixed to
$P=6500$\,d), which is a lower bound to the actual period. The new
preliminary orbital elements obtained by fixing the period to the
above value are listed in Table~\ref{tab6}.

\subsection{Photometrically non-variable S stars}

Among the 16 binary S stars with no photometric variations presented
in this subsection, an orbital solution has been derived for 10 of
them whereas 6 have only minimum-period estimates. The results are
given in Table~\ref{tab7}. The corresponding phase diagrams are
displayed in Fig.~\ref{fig4}.

Note that \bd~$+21^\circ$255 is a visual binary, most probably of
optical nature (Jorissen \& Mayor 1992). Table~\ref{tab7} and
Fig.~\ref{fig4} provide the orbit of the S star (\bd~$+21^\circ$255 =
PPM 91178 = SAO 75009 = HIC 8876), whereas the visual K-type companion
(\bd~$+21^\circ$255p = PPM 91177 = SAO 75008) is also a spectroscopic
binary whose orbit is given in Jorissen \& Mayor (1992).

The jitter level of the non-binary stars of the sample is of the order
of 1-2\kms\ (Jorissen et al. 1998). Its hampering influence on the
detection of binarity is thus limited to low-amplitude orbital
motions.

\section{Mira S stars, SC and C stars: intrinsic radial-velocity variations 
versus orbital motion}
\label{sect5}

\subsection{Intrinsic radial-velocity variations of Mira variables}
\label{sect:intrinsic}

\begin{figure}[th]
\ifigx{8.8 cm}{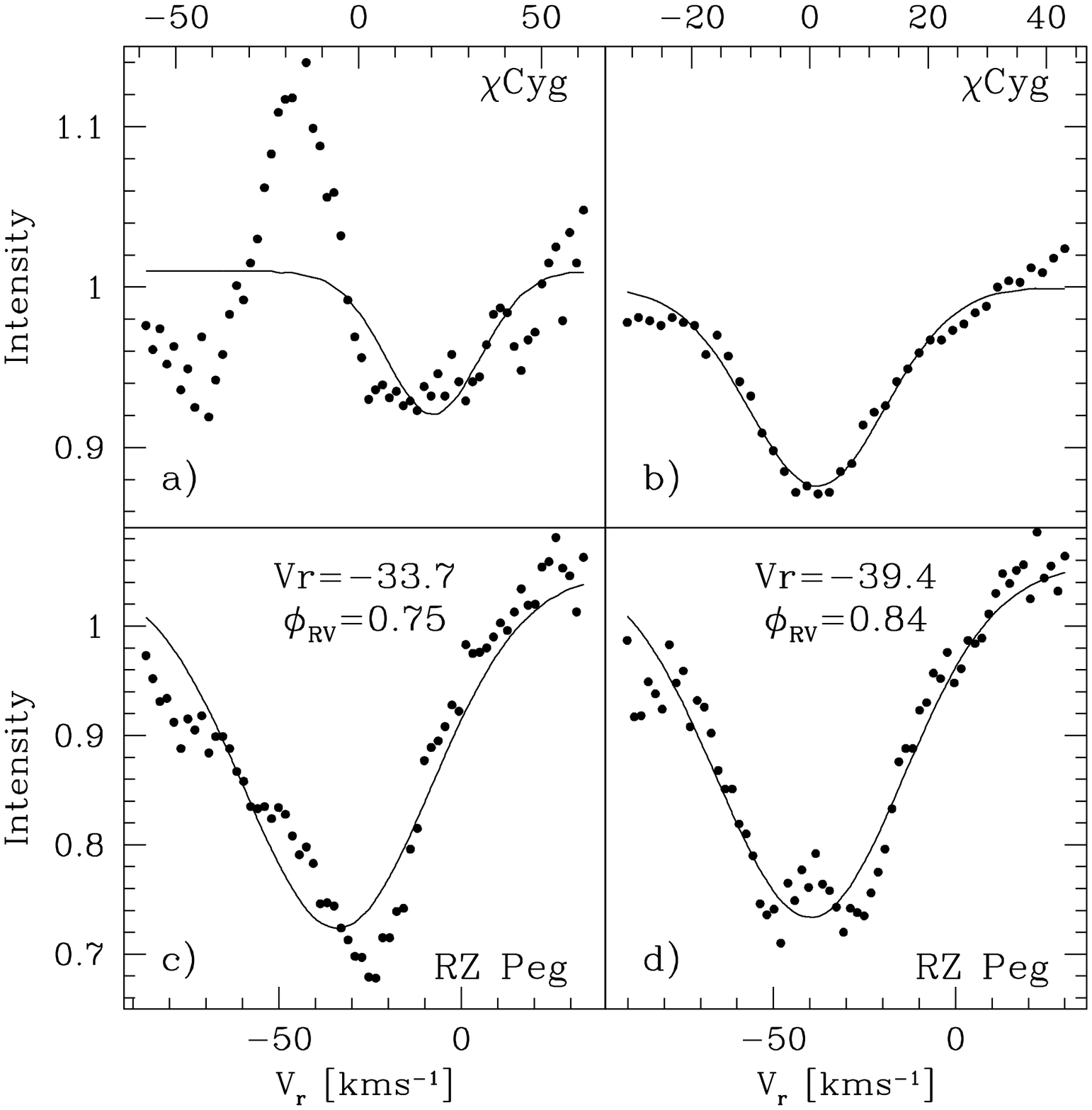}
\caption{Examples of cross-correlation dips for Mira S stars.
a-b)~Profiles for $\chi$~Cyg, including an inverse {\sl P~Cygni-}like
profile;
c-d)~Two profiles at different phases ($\phi_{\rm RV}$) for the star
RZ~Peg. The time-varying weights of the two minima induce a variation
of the measured radial velocity}
\label{fig5}
\end{figure}

The large velocity jitter observed in Mira S stars (Tables 3d, 3e and
4 in Jorissen et al. 1998) is related to their complex and variable
cc-dips (Barbier et al. 1988). In some cases like for the star
$\chi$~Cyg, the cc-dip has an inverse {\sl P-Cygni} shape
(Fig.~\ref{fig5}a), out of which a meaningful radial velocity is very
difficult to extract.  In other cases like AA~Cyg and R~Hya, the
cc-dips are featureless, broad and very stable, and these stars have
the smallest jitter in our sample of Mira S stars (Table 3d of
Jorissen et al. 1998).  For stars like R~And (=\hd~1967) and RZ~Peg
(=\hd~209890), two clearly distinct dips are present.  In RZ Peg for
example (Figs.~\ref{fig5}c-d), the two dips correspond to velocities
of $-25$ and $-50$\,\kms, whereas the center of mass of the star (as
probed by submm observations of the circumstellar CO rotational lines)
moves with a velocity of $-35$\,\kms\ (Sahai \& Liechti 1995), close
to the average of the two \coravel\ dips. The two minima observed in
the cc-dip are therefore likely associated with upwards- and
downwards-moving layers forming a shock in the Mira atmosphere
(e.g. Fox et al. 1984, Querci 1986, Bowen 1988, Bessell et al. 1996).
The respective weights of the two minima vary with time, and so does
the velocity of the resulting blend to which a gaussian is fitted. As
an illustration, 2 profiles at different phases ($\phi_{\rm RV}$) are
given in Figs.~\ref{fig5}c-d for RZ~Peg. The derived {\it mean} radial
velocities are indicated on the figures.

\begin{figure*}[th!]
\ifigx{17.25cm}{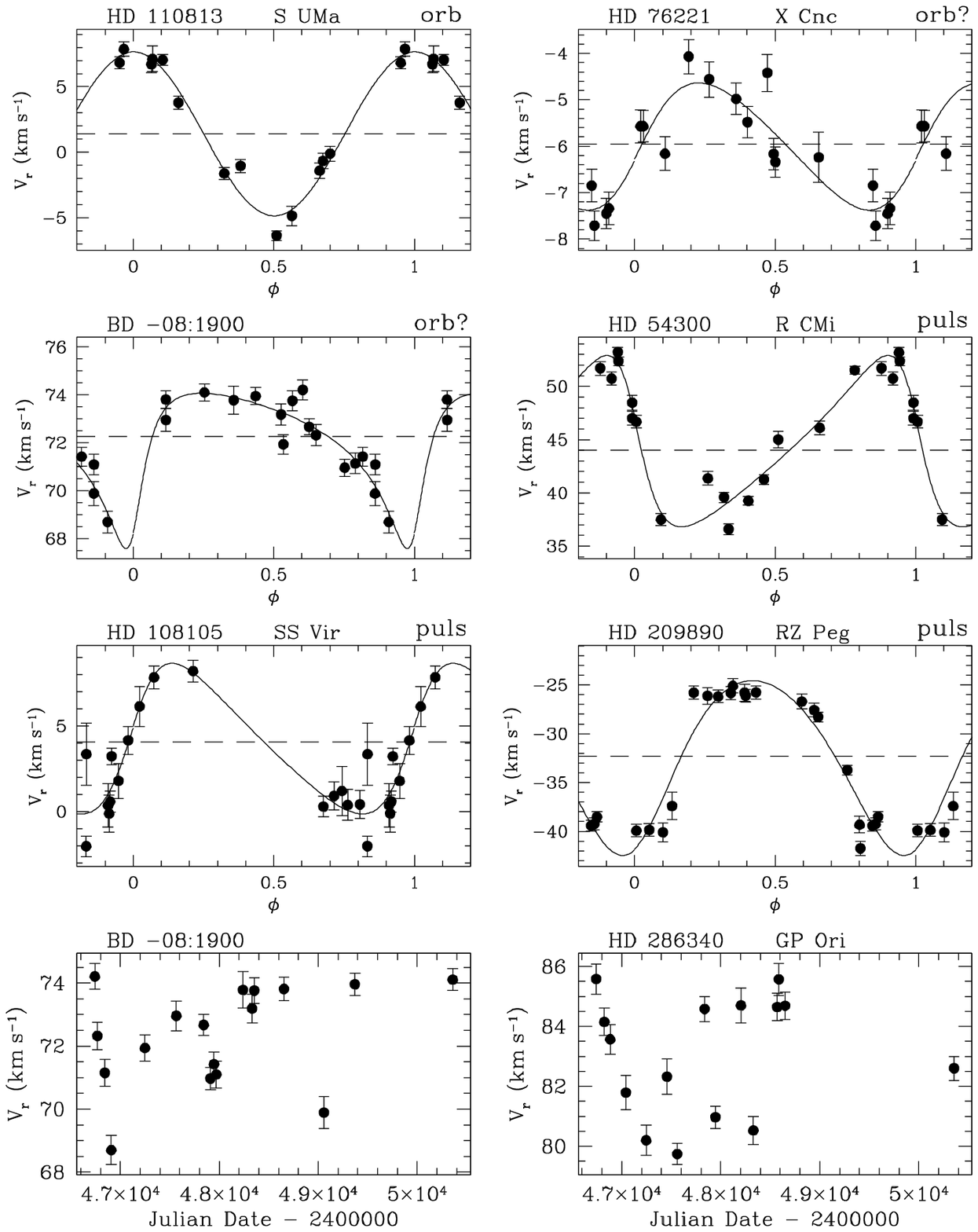}
\caption{Phase-folded radial-velocity curves of Mira, SC and S stars
with {\sl tentative-} or {\sl pseudo-}orbital solutions. Labels
(`orb', `orb?'  or `puls') are used to characterize our confidence
level in the obtained solution or to indicate intrinsic
radial-velocity variations. Two stars (\bd~$-08^\circ 1900$ and
\hd\,286340) have also their radial velocities displayed as a
function of Julian dates}
\label{fig6}
\end{figure*}

When the \coravel\ cc-dip of RZ~Peg is fitted by a {\it single}
gaussian profile, as indicated on Fig.~\ref{fig5}, a {\sl
pseudo-}orbital solution is found (Table~\ref{tab8}), with a period of
437.3$\pm3.9$\,d.  {\it This radial-velocity period may be identified
with the photometric period of 437.8\,d derived from the AAVSO light
curve for the Mira variations} (see Sect.~\ref{sect:RZPeg}). It is
therefore clear that, although they mimick an orbital motion, the
radial-velocity variations of RZ~Peg are intrinsic to the Mira
phenomenon. They will be discussed in more details in
Sect.~\ref{sect:RZPeg}. The same coincidence between the periods of
the light variations and of the velocity variations is obtained for
the CS Mira star R~CMi (=\hd\,54300) and for the C
semi-regular (SRa) star SS~Vir (=\hd~108105) as well. The
radial-velocity variations of R~CMi can be fitted with a {\sl
pseudo-}orbital solution of period 337.3$\pm1.3$\,d, whereas the GCVS
lists $P = 337.8$\,d for the associated Mira variations. It is
interesting to note that the radial-velocity variations of R~CMi are
strongly non-sinusoidal in spite of the almost symmetric light curve
(the rise time from minimum to maximum light represents 48\% of the
total cycle, according to the GCVS, compared to 44\% for RZ Peg). For
SS~Vir, the radial-velocity and photometric periods are
361.2$\pm3.4$\,d and $\sim 364$\,d, respectively.  The relevant
parameters of the {\sl pseudo-}orbits for RZ~Peg, R~CMi and SS~Vir are
given in Table~\ref{tab8}. The radial-velocity periods have been used
to fold the points in phase. The corresponding diagrams are displayed
in Fig.~\ref{fig6}.

\subsection{Intrinsic radial-velocity variations of Mira S stars:
the cases of RZ~Peg and $\chi$~Cyg}
\label{sect:RZPeg}

The radial-velocity curve of RZ~Peg is well sampled, and makes it
possible to investigate in more details the properties of the double
cc-dip phenomenon.

\begin{figure}[t]
\ifigx{8.8 cm}{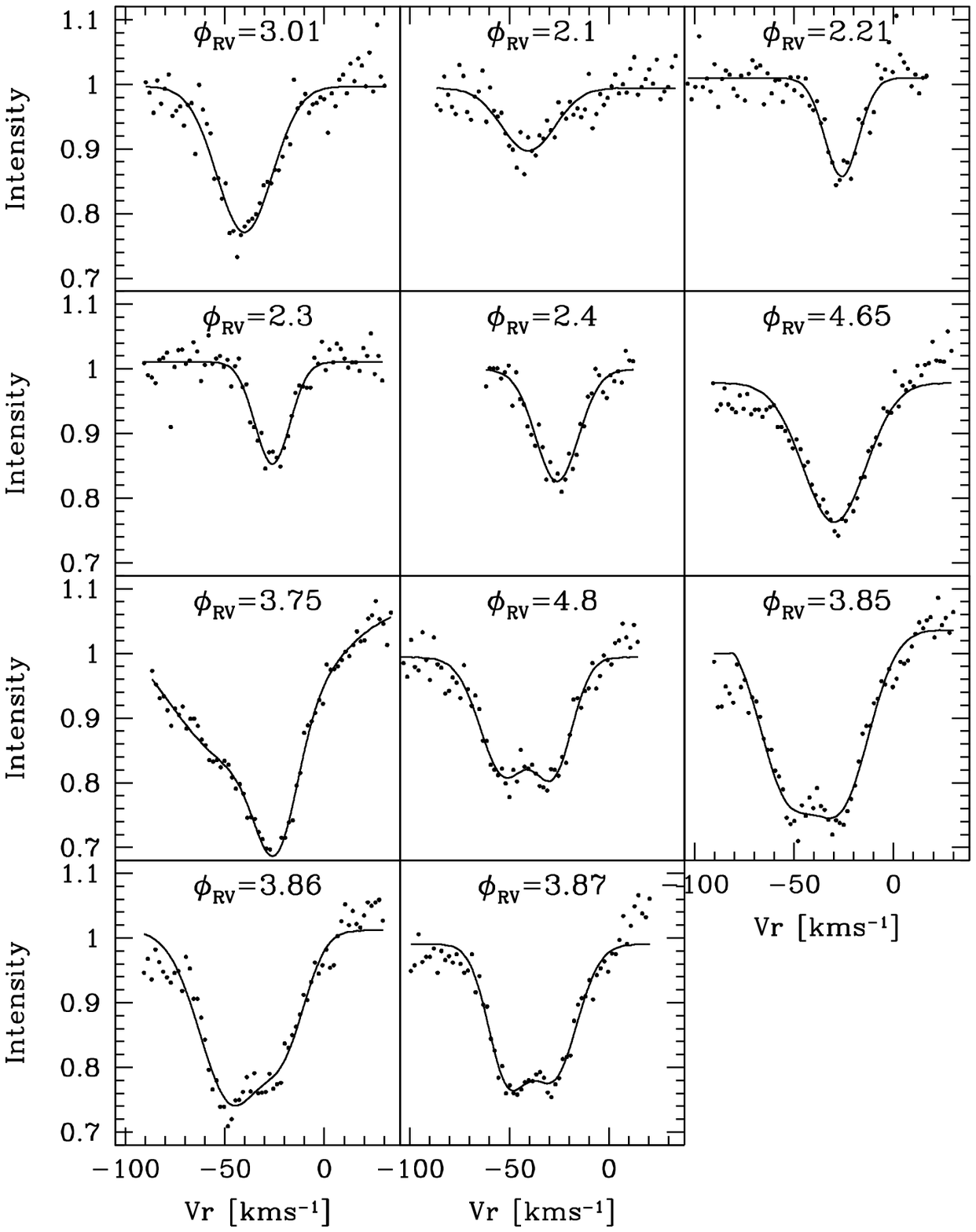}
\caption{\coravel\ cc-dips of RZ~Peg as a function of the
radial-velocity phase ($\phi_{\rm RV}$). Phase 0 (HJD 2\ts448\ts413.3)
is close to the epoch of minimum radial-velocity
(Fig.~\protect\ref{fig6}). The profiles have been fitted with a single
or double gaussian function, when appropriate (thin solid line)}
\label{fig7}
\end{figure}

\begin{figure}
\ifigx{8.8 cm}{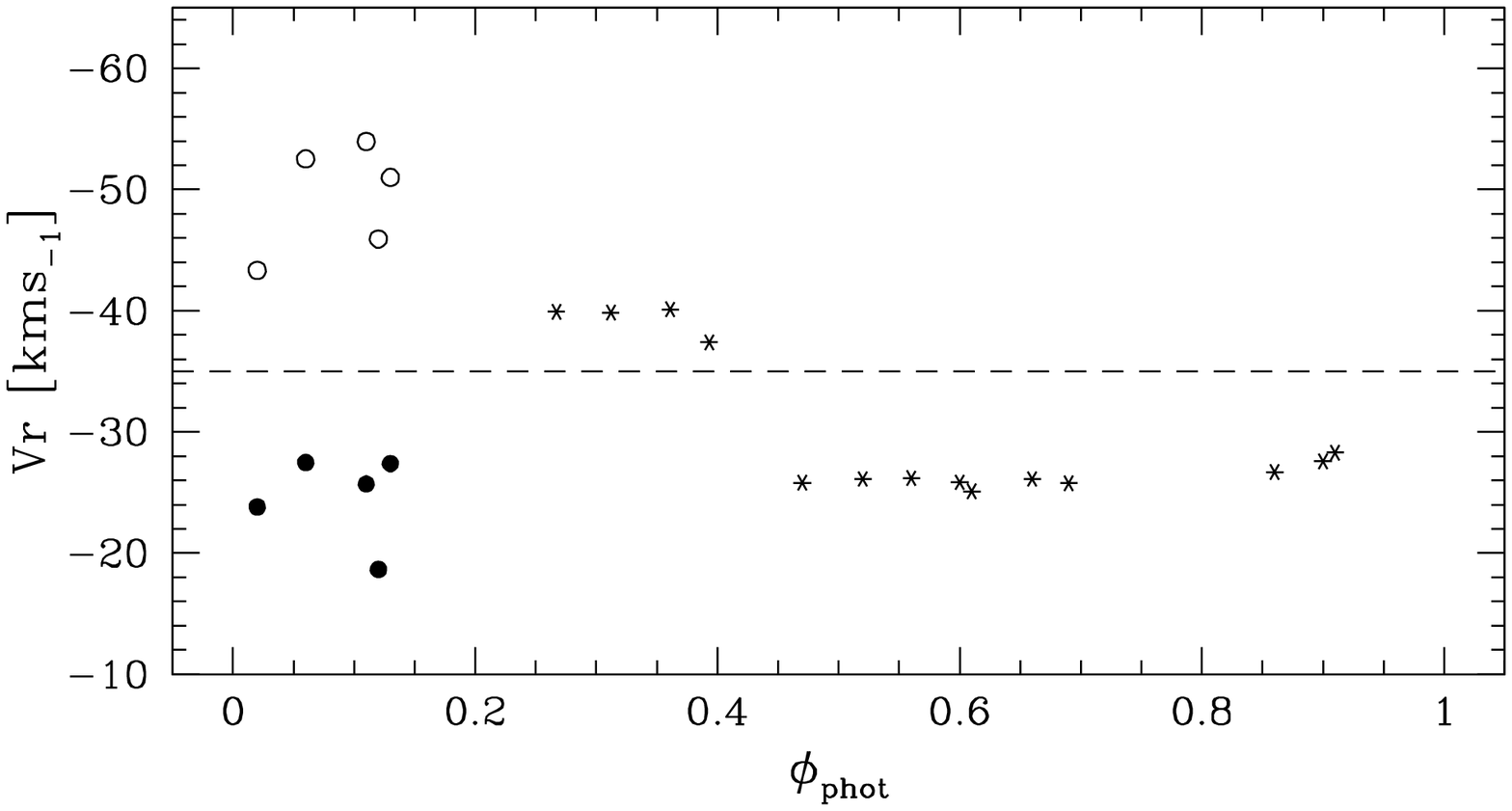}
\caption{Radial velocities of RZ~Peg as a function of the photometric
phase ($\phi_{phot}=\phi_{RV}+0.26$).  Open and filled circles
indicate the velocities of the double-line reductions whereas
star-like symbols are used for velocities derived from gaussian
one-line fitting. The center-of-mass velocity, as probed by submm
observations of the CO rotational lines (Sahai \& Liechti 1995), is
represented by the horizontal dashed line}
\label{fig8}
\end{figure}

Figure~\ref{fig7} presents the evolution of the \coravel\
cross-correlation profiles as a function of the radial-velocity phase
(derived from the elements listed in Table~\ref{tab8}).  The evolution
of the profiles, from a single cc-dip to a double cc-dip and {\it vice
versa}, is quite obvious. The radial velocities corresponding to each
of the peaks have been extracted by fitting a double-gaussian function
to the profile. The resulting velocity curve as a function of the
phase is displayed in Fig.~\ref{fig8}. Two conclusions may be drawn
from this figure: (i) each peak has a rather constant velocity, with
the center-of-mass velocity corresponding almost to the average of the
two peaks, and (ii) the double dip occurs around velocity phases
$\phi_{\rm RV} = 0.7$ -- 0.9, and this behaviour repeats from one
cycle to the other. The velocity difference between the two peaks is
of the order of 20 -- 30~\kms.

The light curve of RZ~Peg has been kindly provided to us by the AAVSO
(Mattei 1997), and makes it possible to convert velocity phases into
photometric phases.  The photometric period appears reasonably stable
over the time span covered by the radial-velocity observations, with
an average period of 437.8\,d, very close to the 437.3\,d period of
the velocity variations.  Adopting a period of 437.3\,d, the relation
$\phi_{\rm phot} = \phi_{\rm RV} + 0.26$ is derived from the AAVSO
maximum (HJD 2\ts448\ts301) closest to the zero point of the
radial-velocity phase (HJD 2\ts448\ts413).  From this relation, it may
be concluded that {\em the double cc-dip occurs just after maximum
light} ($\phi_{\rm phot} = 0.0 - 0.2$).

A similar phenomenon of line doubling near maximum light has been
reported in several other Mira stars.
In the Mira S star $\chi$~Cyg, Hinkle et al. (1982) reported such a
line doubling for the rotation-vibration second overtone ($\Delta v =
3$) infrared CO lines and high-excitation first overtone ($\Delta v =
2$) CO lines. Maehara (1971) also reported the doubling of CN and
atomic lines near 800.0\,nm around maximum light in $\chi$~Cyg, and
Gillet et al.  (1985) in S Car. This line-doubling phenomenon is the
signature of a shock front propagating through the line-forming region
(e.g. Gillet et al. 1985, Bessell et al. 1996).  Incidentally, in
$\chi$~Cyg (Fig.~\ref{fig5}), a line-doubling as clear as in RZ~Peg is
{\it not} observed for the blue-violet iron lines sampled by \coravel\
(Baranne et al. 1979). The \coravel\ radial velocities of $\chi$~Cyg
(averaging to +3.2~\kms; Table~3d of Jorissen et al. 1998), derived
from the clean cc-dips (Fig.~5b), agree with the value obtained by
Pierce et al. (1979).  Hinkle et al. (1982; their Fig.~15) argue that
these blue-violet photospheric absorption lines with a constant radial
velocity form in the infalling material for the whole pulsation cycle.
Nevertheless, the \coravel\ profile of $\chi$~Cyg exhibits some
characteristic features that may be related to the shock front
propagating through the photosphere.  Near visual phases 0.0--0.2, the
\coravel\ profile of $\chi$~Cyg has a typical inverse {\sl P-Cygni}
shape (Fig.~5a) with the blue-shifted `emission' component peaking at
$-20$~\kms. This velocity corresponds to that of the outflowing
material, as derived from the second overtone CO rotation-vibration
lines by Hinkle et al. (1982). A similar inverse P-Cygni profile has
been reported by Ferlet \& Gillet (1984) for the TiI lines near 1
$\mu$m in Mira near maximum light.

It is very likely that the same physical phenomenon -- namely a shock
front propagating through the line-forming region -- is responsible
for the time-dependent features observed in the \coravel\ cc-dips of
RZ~Peg and $\chi$~Cyg. This conclusion is suggested by the fact that
in both stars, the line-doubling occurs near maximum light, and the
offset between the two distinct components of the cc-dip is of the
same order (20 to 30~\kms, corresponding to the shock
discontinuity). These observed features are well reproduced by the
synthetic FeI and CO line profiles computed by Bessell et al. (1996)
in a dynamical Mira atmosphere with a propagating shock.  For some
reason however, the atmospheric structures of RZ~Peg and $\chi$~Cyg
must be different, so that only in RZ~Peg are the blue-violet iron
lines sampled by \coravel\ forming in absorption on both sides of the
shock, resulting in a clean double-minima cc-dip.

\subsection{Binaries among Mira variables and SC stars}
\label{sect:binaryMira}

Because they are difficult to observe at minimum light, Mira S and C
stars were generally insufficiently sampled to attempt a period search
on their radial-velocity curve. Well-sampled curves are, however,
available for the Mira S stars AA~Cyg, $\chi$~Cyg, R~Hya, for the C
(no Tc) stars SS~Vir and S~UMa, and for all the SC stars. Some of
these stars have already been discussed in Sects.~\ref{sect:intrinsic}
and \ref{sect:RZPeg}.
  
No satisfactory periods emerge for AA Cyg and R~Hya. 
The radial velocity of the absorption lines in $\chi$~Cyg is constant,
as discussed in Sect.~\ref{sect:RZPeg}.  The Mira S star S~UMa
(=\hd~110813) is perhaps a binary, since the radial-velocity period $P
= 592.2$\,d (Table~\ref{tab8}) is well distinct from the 225.9\,d
period of the Mira variations.  More measurements are needed, however,
before that orbital solution may definitely be accepted.

As far as Tc-poor carbon stars are concerned, X Cnc (= \hd~76221) is
probably binary, with a 491\,d orbit (quite distinct from the
-uncertain- 195\,d period quoted by the GCVS for the SRb variations).

A possible orbital solution, with $P = 544.2\pm5.7$\,d and $e = 0.55$,
has been found for the SC star \bd~$-08^\circ1900$ (Table~\ref{tab8}
and Fig.~\ref{fig6}). A light curve for this star is provided by
Jorissen et al. (1997), who find photometric variations on a time
scale of about 59\,d. The absence of coincidence between the
radial-velocity and photometric (qua\-si-)periods may lend some credit
to the orbital nature of the radial-velocity variations observed for
\bd~$-08^\circ1900$.  We do not accept this interpretation, however,
without the following word of caution: another SC star, GP~Ori
(=\hd\,286340), exhibits radial-velocity variations of a nature
very similar to those of \bd~$-08^\circ1900$ (Fig.~\ref{fig6}; note
especially the drift observed in the early phase of the monitoring,
reminiscent of that observed for \bd~$-08^\circ1900$), but no orbital
solution could be obtained for GP~Ori.  At this point, we cannot
exclude the possibility that the `orbital' solution found for
\bd~$-08^\circ1900$ is simply a consequence of a favourable time
sampling of irregular, intrinsic variations. One should note in that
respect that the $\sigma_{\rm V}/\sigma_{\rm Vr}$ ratio observed for
\bd~$-08^\circ1900$ fits well the predictions of a simple linear model
of adiabatic acoustic oscillations (Jorissen et al. 1997).  Finally,
the orbital parameters of \bd~$-08^\circ1900$ would locate that star
in a unusual region of the $(e, \log P)$ diagram (Fig.~4 of Jorissen
et al. 1998), which is another argument against that orbital
solution.

In fact, relatively large-amplitude radial-velocity variations
($\sigma_{\rm Vr}$ of a few \kms; Table 3e of Jorissen et al.  1998)
are a common feature among SC stars.  For the two Mira SC stars in our
sample (R~CMi and RZ~Peg), periodic radial-velocity variations were
found with the same period as the light curve
(Sect.~\ref{sect:intrinsic}).  For the remaining SC stars, which are
of semiregular (SR) or irregular (L) variability types, very irregular
radial-velocity variations are indeed observed, with the possible
exception of \bd~$-08^\circ1900$ discussed above.  When the number of
measurements is relatively small ($<10$) and covers a limited time
span, these variations may mimick orbital variations.

\acknowledgements {We thank the AAVSO for communicating us the light curve of
RZ Peg. This research has been supported partly by the {\it Fonds
National de la Recherche Scientifique} (Switzerland, Belgium) and the
University of Geneva (Geneva Observatory).  S.V.E. is supported by a
F.R.I.A. doctoral grant (Belgium). A.J.  is Research Associate
(F.N.R.S., Belgium).}

\end{document}